\documentclass[epj,final]{svjour}
\usepackage{epsfig, graphicx}
\begin{document}

\title{Baryon resonance production and dielectron decays in proton-proton collisions at 3.5~GeV}

\author{G.~Agakishiev$^{7}$, A.~Balanda$^{3}$, D.~Belver$^{18}$, A.~Belyaev$^{7}$,
J.C.~Berger-Chen$^{9}$, A.~Blanco$^{2}$, M.~B\"{o}hmer$^{10}$, J.~L.~Boyard$^{16}$, P.~Cabanelas$^{18}$,
S.~Chernenko$^{7}$, A.~Dybczak$^{3,*}$, E.~Epple$^{9}$, L.~Fabbietti$^{9}$, O.~Fateev$^{7}$,
P.~Finocchiaro$^{1}$, P.~Fonte$^{2,b}$, J.~Friese$^{10}$, I.~Fr\"{o}hlich$^{8}$, T.~Galatyuk$^{5,c}$,
J.~A.~Garz\'{o}n$^{18}$, R.~Gernh\"{a}user$^{10}$, K.~G\"{o}bel$^{8}$, M.~Golubeva$^{13}$, D.~Gonz\'{a}lez-D\'{\i}az$^{5}$,
F.~Guber$^{13}$, M.~Gumberidze$^{5,c}$, T.~Heinz$^{4}$, T.~Hennino$^{16}$, R.~Holzmann$^{4}$,
A.~Ierusalimov$^{7}$, I.~Iori$^{12,e}$, A.~Ivashkin$^{13}$, M.~Jurkovic$^{10}$, B.~K\"{a}mpfer$^{6,d}$,
T.~Karavicheva$^{13}$, I.~Koenig$^{4}$, W.~Koenig$^{4}$, B.~W.~Kolb$^{4}$, G.~Kornakov$^{5}$,
R.~Kotte$^{6}$, A.~Kr\'{a}sa$^{17}$, F.~Krizek$^{17}$, R.~Kr\"{u}cken$^{10}$, H.~Kuc$^{3,16}$,
W.~K\"{u}hn$^{11}$, A.~Kugler$^{17}$, A.~Kurepin$^{13}$, V.~Ladygin$^{7}$, R.~Lalik$^{9}$,
S.~Lang$^{4}$, K.~Lapidus$^{9}$, A.~Lebedev$^{14}$, T.~Liu$^{16}$, L.~Lopes$^{2}$,
M.~Lorenz$^{8,c}$, L.~Maier$^{10}$, A.~Mangiarotti$^{2}$, J.~Markert$^{8}$, V.~Metag$^{11}$,
B.~Michalska$^{3}$, J.~Michel$^{8}$, C.~M\"{u}ntz$^{8}$, L.~Naumann$^{6}$, Y.~C.~Pachmayer$^{8}$,
M.~Palka$^{3}$, Y.~Parpottas$^{15,f}$, V.~Pechenov$^{4}$, O.~Pechenova$^{8}$, J.~Pietraszko$^{4}$,
W.~Przygoda$^{3,*}$, B.~Ramstein$^{16}$, A.~Reshetin$^{13}$, A.~Rustamov$^{8}$, A.~Sadovsky$^{13}$,
P.~Salabura$^{3}$, A.~Schmah$^{a}$, E.~Schwab$^{4}$, J.~Siebenson$^{9}$, Yu.G.~Sobolev$^{17}$,
S.~Spataro$^{g}$, B.~Spruck$^{11}$, H.~Str\"{o}bele$^{8}$, J.~Stroth$^{8,4}$, C.~Sturm$^{4}$,
A.~Tarantola$^{8}$, K.~Teilab$^{8}$, P.~Tlusty$^{17}$, M.~Traxler$^{4}$, R.~Trebacz$^{3}$,
H.~Tsertos$^{15}$, T.~~Vasiliev$^{7}$, V.~Wagner$^{17}$, M.~Weber$^{10}$, C.~Wendisch$^{6,d}$,
J.~W\"{u}stenfeld$^{6}$, S.~Yurevich$^{4}$, Y.~Zanevsky$^{7}$}

\institute{Istituto Nazionale di Fisica Nucleare - Laboratori Nazionali del Sud, 95125~Catania, Italy \and
LIP-Laborat\'{o}rio de Instrumenta\c{c}\~{a}o e F\'{\i}sica Experimental de Part\'{\i}culas , 3004-516~Coimbra, Portugal \and
Smoluchowski Institute of Physics, Jagiellonian University of Cracow, 30-059~Krak\'{o}w, Poland \and
GSI Helmholtzzentrum f\"{u}r Schwerionenforschung GmbH, 64291~Darmstadt, Germany \and
Technische Universit\"{a}t Darmstadt, 64289~Darmstadt, Germany \and
Institut f\"{u}r Strahlenphysik, Helmholtz-Zentrum Dresden-Rossendorf, 01314~Dresden, Germany \and
Joint Institute of Nuclear Research, 141980~Dubna, Russia \and
Institut f\"{u}r Kernphysik, Goethe-Universit\"{a}t, 60438 ~Frankfurt, Germany \and
Excellence Cluster 'Origin and Structure of the Universe' , 85748~Garching, Germany \and
Physik Department E12, Technische Universit\"{a}t M\"{u}nchen, 85748~Garching, Germany \and
II.Physikalisches Institut, Justus Liebig Universit\"{a}t Giessen, 35392~Giessen, Germany \and
Istituto Nazionale di Fisica Nucleare, Sezione di Milano, 20133~Milano, Italy \and
Institute for Nuclear Research, Russian Academy of Science, 117312~Moscow, Russia \and
Institute of Theoretical and Experimental Physics, 117218~Moscow, Russia \and
Department of Physics, University of Cyprus, 1678~Nicosia, Cyprus \and
Institut de Physique Nucl\'{e}aire (UMR 8608), CNRS/IN2P3 - Universit\'{e} Paris Sud, F-91406~Orsay Cedex, France \and
Nuclear Physics Institute, Academy of Sciences of Czech Republic, 25068~Rez, Czech Republic \and
LabCAF. F. F\'{\i}sica, Univ. de Santiago de Compostela, 15706~Santiago de Compostela, Spain }

\mail{witold.przygoda@uj.edu.pl}

\date{Received: date / Revised version: date}

\abstract{ We report on baryon resonance production and decay in proton-proton collisions at a kinetic energy of $3.5$ GeV based on data measured with HADES. The exclusive channels $pp \rightarrow np\pi^{+}$ and $pp \rightarrow pp\pi^{0}$ as well as $pp \rightarrow ppe^{+}e^{-}$ are studied simultaneously for the first time. The invariant masses and angular distributions of the pion-nucleon systems were studied and compared to simulations based on a resonance model ansatz assuming saturation of the pion production by an incoherent sum of baryonic  resonances (R) with masses $<2~$ GeV/$c^2$. A very good description of the one-pion production is achieved allowing for an estimate of individual baryon-resonance production-cross-sections which are used as input to calculate the dielectron yields from $R\rightarrow pe^+e^-$ decays. Two models of the resonance decays into dielectrons are examined assuming a point-like $RN \gamma^*$ coupling and the dominance of the $\rho$ meson. The results of model calculations are compared to data from the exclusive $ppe^{+}e^{-}$ channel by means of the dielectron and $pe^+e^-$ invariant mass distributions.
}

\authorrunning{G.~Agakishiev et al.}
\titlerunning{Baryon resonance production and dielectron decays in proton-proton collisions at 3.5~GeV}
\maketitle

\begingroup
\renewcommand{\thefootnote}{\alph{footnote}}
\footnotetext{$^{*}$ corresponding authors}
\endgroup

\section{Introduction}
\label{intro}

The investigation of baryon resonance ($R$) decays into a nucleon ($N$) and a massive (virtual) photon ($\gamma^*$) provides a unique opportunity to explore the resonance structure. It gives complementary information to the one obtained from experiments studying resonance production by means of electron or photon beams. The interaction vertex ($RN\gamma^*$) is described by a set of electromagnetic Transition Form Factors (eTFF), depending on the resonance isospin, spin, parity and the four momentum squared ($q^2$) of the virtual photon. While in the electro-production experiments $q^2<0$, where the respective form factors are accessible in the space-like region, the time-like region ($q^2>0$) can be probed by the process of resonance transition into $N e^+e^-$ (commonly named Dalitz decay). A rich data sample of the transition amplitudes for $\Delta(1232)$, $N(1440)$ and $N(1520)$ has been obtained in the space-like region in a wide $q^2$ range. Comparison of the data to various model calculations allows to estimate contributions originating from a quark core and a pion cloud (for a  review see \cite{jlab}). The latter appears to be particulary important at small $q^2$, contributing significantly to the respective eTFF, as for example shown for the $\Delta(1232)$. On the other hand, no experimental data on the Dalitz decays of resonances exist, though many theoretical calculations predict a sensitivity of the dilepton invariant mass distribution to the $RN\gamma^*$ vertex structure. Indeed, according to the Vector Meson Dominance (VMD) model of Sakurai \cite{sakurai} the virtual photon coupling to a hadron is mediated entirely by intermediate vector mesons $\rho$/$\omega$/$\phi$. Hence, it is expected that the contribution of mesons to the interaction vertex modifies the $q^2$ dependence of the respective eTFF and produces an enhancement near the vector meson poles. However, it has also been realized that such strict VMD leads to an overestimation of the radiative $R\rightarrow N\gamma$ decay widths when the known $R\rightarrow N\rho$ branching ratios are used in calculations (see e.g. \cite{kroll,faessler}). Various solutions of this problem were proposed, as for example the application of two independent coupling constants for the vector mesons and photon \cite{kroll}, destructive interferences between contributions from higher $\rho/\omega$ states \cite{kriv} or different couplings to the quark core and pion cloud \cite{pena}. The salient feature of all these models, however, is a significant modification of the eTFF due to the vector meson-resonance couplings.

\begingroup
\renewcommand{\thefootnote}{\alph{footnote}}
\footnotetext{$^{a}$ also at Lawrence Berkeley National Laboratory, ~Berkeley, USA}
\footnotetext{$^{b}$ also at ISEC Coimbra, ~Coimbra, Portugal}
\footnotetext{$^{c}$ also at ExtreMe Matter Institute EMMI, 64291~Darmstadt, Germany}
\footnotetext{$^{d}$ also at Technische Universit\"{a}t Dresden, 01062~Dresden, Germany}
\footnotetext{$^{e}$ also at Dipartimento di Fisica, Universit\`{a} di Milano, 20133~Milano, Italy}
\footnotetext{$^{f}$ also at Frederick University, 1036~Nicosia, Cyprus}
\footnotetext{$^{g}$ also at Dipartimento di Fisica Generale and INFN, Universit\`{a} di Torino, 10125~Torino, Italy}
\endgroup

Understanding the couplings of vector-meson resonances is of utmost importance also for another but closely connected reason. A strong modification of the $\rho$ meson spectral function is observed in dilepton invariant mass distributions measured in ultra-relativistic heavy ion collisions at SPS \cite{ceres,na60} and also at RHIC \cite{rhic,phenix}. The experimental findings are consistently explained by model calculations assuming strong couplings of the $\rho$ meson to baryon-resonance $-$ nucleon-hole states excited in hot and dense nuclear matter \cite{rap}. Similar calculations for cold nuclear matter predict also strong off-shell $\rho$ couplings to the low-mass baryon resonances like $N(1440)$, $N(1520)$, $N(1720)$ and $\Delta(1620)$ shifting part of the strength of the $\rho$ meson spectral function down below the meson pole \cite{peters} (for recent review see also \cite{metag}). The respective coupling strengths are usually constrained in models by the data from meson photo-production and/or known resonance-$\rho N$ branchings and extrapolations assuming VMD (see for example \cite{wambach}). An independent experimental information, however, would be extremely important for a validation of these calculations. Pion induced reactions, as for example $\pi^-p\rightarrow e^+e^-n$, are ideally suited for such investigation, but have not been studied yet. Alternative reaction channels like proton-proton collisions at low bombarding energies can be used, yet, at the expense of a more complicated description of the resonance production.

To begin with a discussion of proton-proton reactions, we shall recall the results of first high statistics measurements of inclusive $e^+e^-$ production in $p+p$ and $p+Nb$ collisions at $3.5$ GeV kinetic energy \cite{pp35GeV,pNb35GeV}. The comparison of the measured dielectron invariant mass distributions to calculations based on a resonance model \cite{GiBUU} clearly suggests the important role of  $R\rightarrow N\rho\rightarrow Ne^+e^-$ decays. A very good description of the data by the calculation seems to support such a scheme, where dielectrons are produced entirely through the intermediate $\rho$. However, as the authors of \cite{GiBUU} conclude, the obtained results should be treated as an "educated guess" because both resonance production and their dielectron decays are subject to large uncertainties. More exclusive data with various final states are needed to pin down the mechanism of the resonance production and decay. Moreover, in the calculations, a good description of the $e^+e^-$ invariant mass distributions could also be achieved assuming a mass dependent eTFF of the $\Delta(1232)$ \cite{iachello,GiBUU} but neglecting contributions from higher mass baryonic resonances. On the other hand, such strong modification of the $\Delta(1232)$ eTFF leads to an overestimate of the dielectron yield at high transverse momentum and is not confirmed by recent calculations \cite{pena}.

The GiBUU model \cite{GiBUU,GiBUU1} uses a  parametrization of the resonance production cross sections according to the model of Teis et al. \cite{teis}. This model assumes constant matrix elements for the resonance production, except the $\Delta(1232)$, where the results of a One-Pion-Exchange (OPE) calculation \cite{dmitriev} are adopted. In our earlier studies of one-pion production in $p+p$ reactions at $1.25$ and $2.2$ GeV we have shown that this model describes the data well if the angular distributions of the dominant $\Delta(1232)$ are slightly modified with respect to the original OPE results \cite{pp2GeV}. There are, however, also other prescriptions to parameterize resonance production amplitudes, as for example the one used in the UrQMD transport model \cite{urqmd}. Although the corresponding calculations overestimate the inclusive $e^+e^-$ production in $p+p$ at $3.5$ GeV \cite{pp35GeV}, a more detailed comparison to exclusive data on one-pion and dielectron production is necessary to conclude on the reason of the discrepancy. On the other hand, there are also calculations based on the Lund string model \cite{GiBUU2,hsd} which include explicitly solely $\Delta(1232)$ resonance production and model the vector meson production via string fragmentation. The latter also predicts a very different shape and yield of the dielectron invariant mass distribution resulting from $\rho$ meson decays. Therefore, exclusive data are necessary to clarify the question about resonance production and their contribution to dielectron production in this energy range. The investigations are also important for the future HADES and Compressed Baryonic Matter (CBM) programs at FAIR which address studies of dielectron production in the $3-10$ AGeV beam energy range.

In this work, we present results from three exclusive channels: $pp\rightarrow pn\pi^+$, $pp\rightarrow pp\pi^0$ and $pp\rightarrow ppe^+e^-$ investigated at the kinetic beam energy of $3.5$ GeV ($\sqrt{s}=3.18$ GeV in our fixed-target experiment). The analysis of the first two channels is focused on one-pion production with the aim to learn about the baryon resonance excitation. We show a detailed comparison to simulations based on the resonance model \cite{teis} and determine baryon resonance production cross sections. The obtained cross sections are used to calculate dielectron Dalitz yields which are compared to the ones measured in the exclusive $ppe^+e^-$ channel. Such channel selects, from many other possible dielectron sources, only those which are related to the two-body vector meson decays and the resonance conversions $R\rightarrow pe^+e^-$. The other dielectron sources dominating the inclusive $e^+e^-$ production, in particular the Dalitz decays of $\eta(\pi^0)\rightarrow e^+e^-\gamma$ and $\omega\rightarrow \pi^0e^+e^-$, can be effectively suppressed via kinematical constraints. In the calculations of the resonance Dalitz decay spectra we use a point like $RN\gamma*$ coupling (constant eTFFs), constrained by experimental data on $R\rightarrow N\gamma$ transitions as given in \cite{zetenyi}. We are going to show that modifications of the respective eTFF due to the resonance-vector meson couplings will be directly visible in the $e^+e^-$ invariant mass distributions. In the next steps we compare then the exclusive $ppe^+e^-$ data to the calculations assuming dominance of the $\rho$ meson.

Our work is organized as follows. In Section \ref{exp} we present experimental conditions, apparatus and principles of the particle identification and momentum reconstruction. We also explain the methods used to separate the exclusive reaction channels and to normalize the experimental yields. In Section \ref{sim} we discuss our simulation chain consisting of the event generator and model of the detectors, which is used to determine its acceptance and the reconstruction efficiency. In Section \ref{final states} we present our results on the hadronic $pn\pi^+$ and $pp\pi^0$ final states, and in Section \ref{ppee} we discuss the $ppe^+e^-$ final state and comparisons to the above mentioned models. We close with conclusions and outlook in subection \ref{outlook}.

\section{Experiment}
\label{exp}
\subsection{Detector overview}

The High Acceptance Di-Electron Spectrometer (HADES) consists of six identical sectors covering polar angles $18^0$- $85^0$ with respect to the beam axis. In the experiment a proton beam with intensities of up to $10^7$ particles/s was impinging on a $5$ cm long liquid-hydrogen target ($1\%$ interaction probability). The momentum vectors of produced particles are reconstructed by means of the four drift chambers (MDC) placed before (two) and behind (two) the magnetic field region provided by six coils of a super-conducting toroid. The experimental momentum resolution typically amounts to $2-3\%$ for protons and pions and $1-2\%$ for electrons, depending on the momentum and the polar emission angle. Particle identification (electron/ pion/proton) is provided by a hadron blind Ring Imaging Cherenkov (RICH) detector, centered around the target, two time-of-flight walls based on plastic scintillators covering polar angles larger (TOF) and smaller (TOFINO) than $45^0$, respectively, and a Pre-Shower detector placed behind TOFINO. A detailed description of the spectrometer, track reconstruction and particle identification methods can be found in \cite{hadesspec}.

In the experiment a two-stage hardware trigger was used: (i) the first-level trigger (LVL1) based on hit multiplicity measurements in the TOF/TOFINO walls and (ii) the second-level trigger (LVL2) for electron identification requesting at least one ring in RICH correlated with a fast particle hit in TOF or an electromagnetic cascade in the Pre-Shower detector. The analysis of hadronic channels was based on LVL1 triggered events selected by a hit multiplicity $MUL \ge 3$ in the time-of-flight detectors. The events used for the dielectron analysis were selected using the LVL1 condition and, in addition, a positive LVL2 decision.  All events with a positive LVL2 trigger decision and every third LVL1 event, irrespective of the LVL2 decision, were recorded, yielding a total of $1.17\times 10^9$ events of the reaction p(3.5~GeV)+p.


\subsection{Selection of reaction channels}
\label{channel_selection}

In this work we present results for three exclusive final states: $pp\pi^0$, $pn\pi^+$ and $ppe^+e^-$. The analysis methods are similar to those presented already in detail in \cite{pp2GeV} on $p+p$ collisions at lower beam energies. Below we summarize the most important steps relevant for the analysis presented in this paper.

The channels with pions were selected using events containing at least two tracks from positively charged particles. Particle identification (PID) of the tracks was achieved by the application of two-dimensional selection criteria on the correlation between the velocity ($\beta=v/c$) and the momentum reconstructed in the TOF/TOFINO detectors and the MDC, respectively. Since there was no dedicated start detector in the experiment, a special time of flight reconstruction method was applied, as described below. For each event two hypotheses were tested assuming (i) detection of two protons ($2p$ events) and (ii) detection of one pion and one proton ($p\pi^+$ events). For each hypothesis, both hadrons were considered as reference particles of known masses and momenta. Consequently, the time-of-flight of the reference particle was calculated, and the velocities of all the other reaction products were deduced using only the time-of-flight differences to the reference particles. If there were more than two tracks per event, the procedure was repeated for all two-track combinations and the best was selected by means of a $\chi^2$ test.

For the $ppe^+e^-$ final state, events containing at least one hadron track from a positively charged particle and one dielectron pair were selected. The electron tracks were identified by means of the RICH detector, providing also the electron emission angles for matching with tracks reconstructed in the MDC. In the next step, the event hypothesis  method, described above, was used for all $pe^+e^-$ candidates in a given event. Furthermore, the same procedure was also applied for the $pe^-e^-$ and the $pe^+e^+$ track combinations in order to estimate the combinatorial background (CB) originating mainly from multi-pion ($\pi^{0}$) production followed by a photon conversion in the detector material. The CB was estimated using the like-sign pair technique (given as a sum of like-sign pairs in events with one proton at least), as described in \cite{pp2GeV,hadesspec}.

Finally, the missing masses of two-particle $pp$ and $p\pi^+$ systems, and three-particle $pee$ (for the like-sign and the unlike-sign pairs) systems with respect to the beam-target system were evaluated for a selection of the channels. The subsequent final states were identified via cuts in the one-dimensional missing mass distributions around the value of the not detected particle, $\pi^0$, neutron or proton, respectively. The momentum vectors of not detected particles were obtained from momentum conservation.

\begin{figure}

\hspace{0mm}
\resizebox{0.5\textwidth}{0.22\textheight}{
  \includegraphics{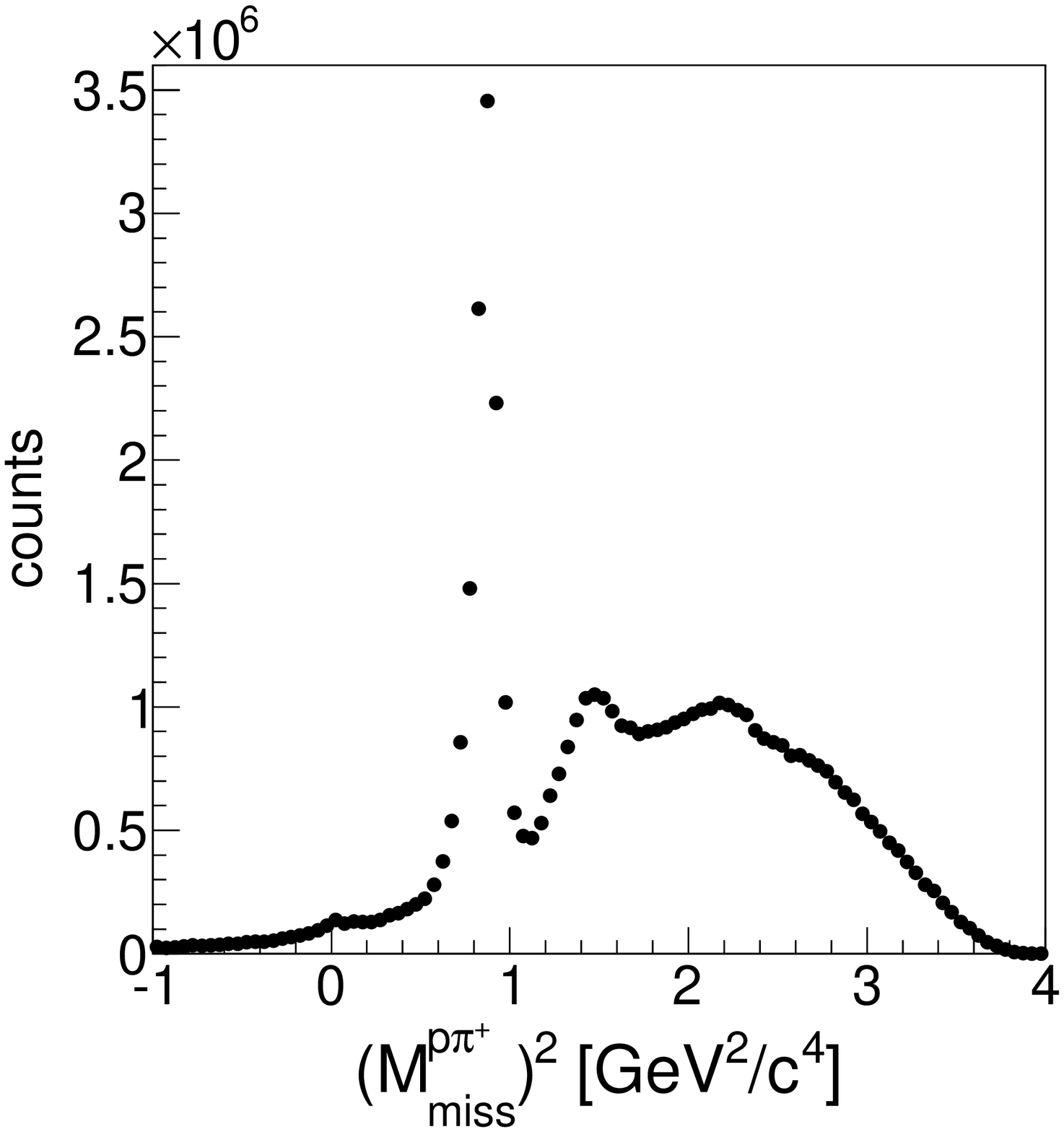}
  \includegraphics{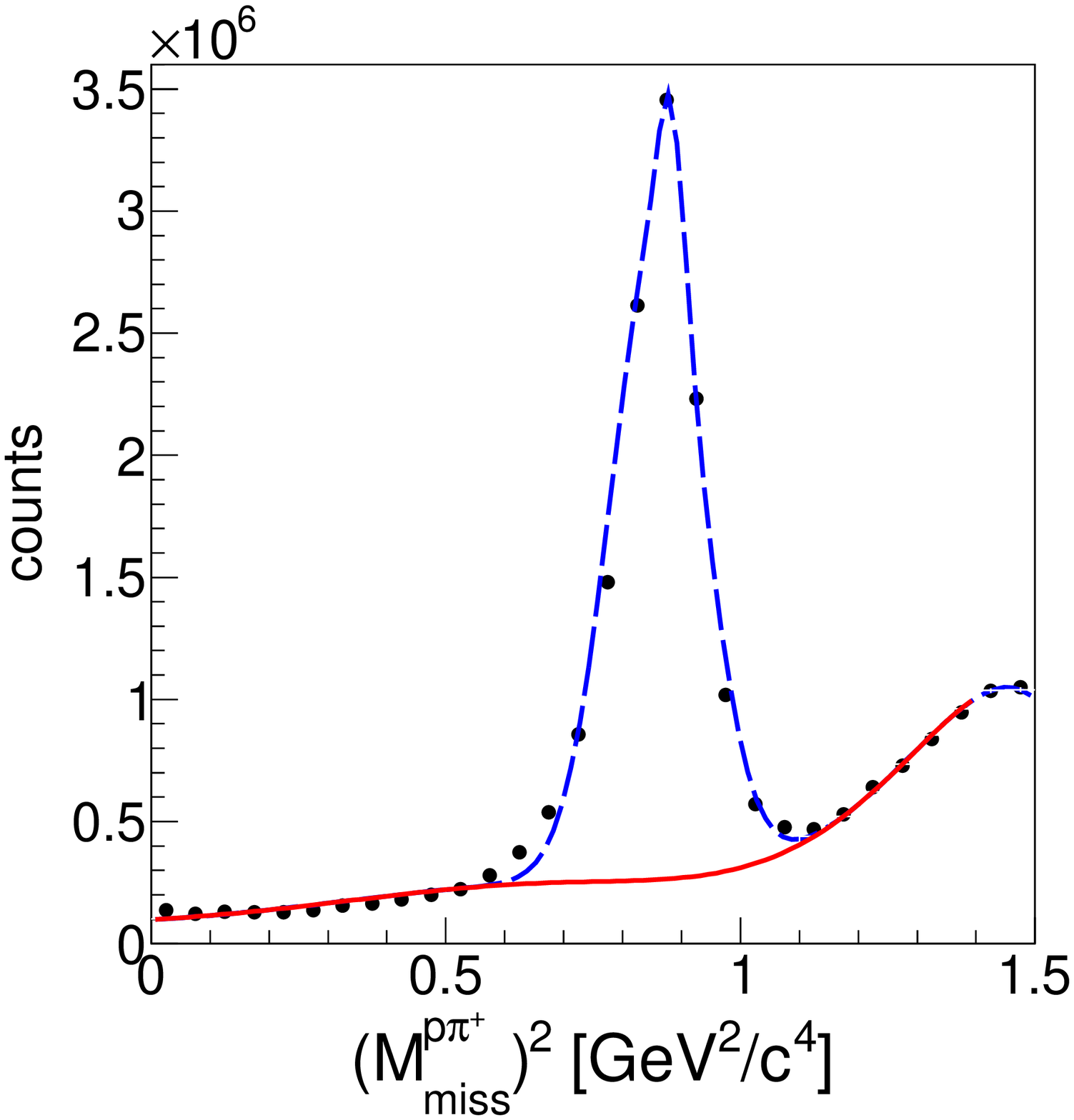}
}

\caption{Left: missing mass squared of the $p\pi^+$ system with respect to the beam-target $pp$ system. Right: an example of a fit within the squared missing mass window around the neutron peak at $(M^{p\pi^+}_{miss})^2$=$0.88~$GeV$^2/c^4$.}
\label{ppip_select}       
\end{figure}

\begin{figure}

\vspace{0mm}
\resizebox{0.5\textwidth}{0.22\textheight}{
  \includegraphics{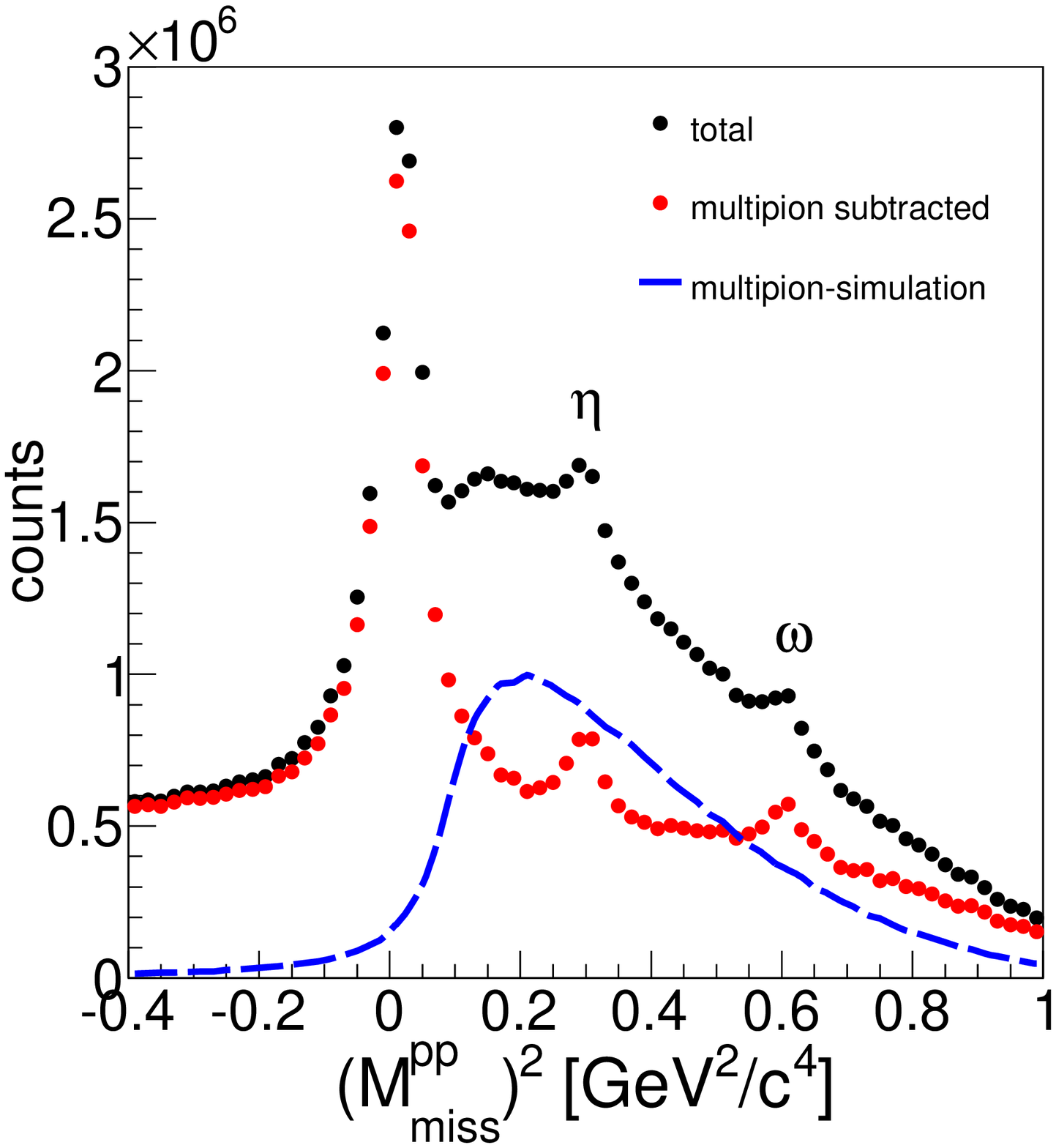}
	\hspace{-15mm}
  \includegraphics{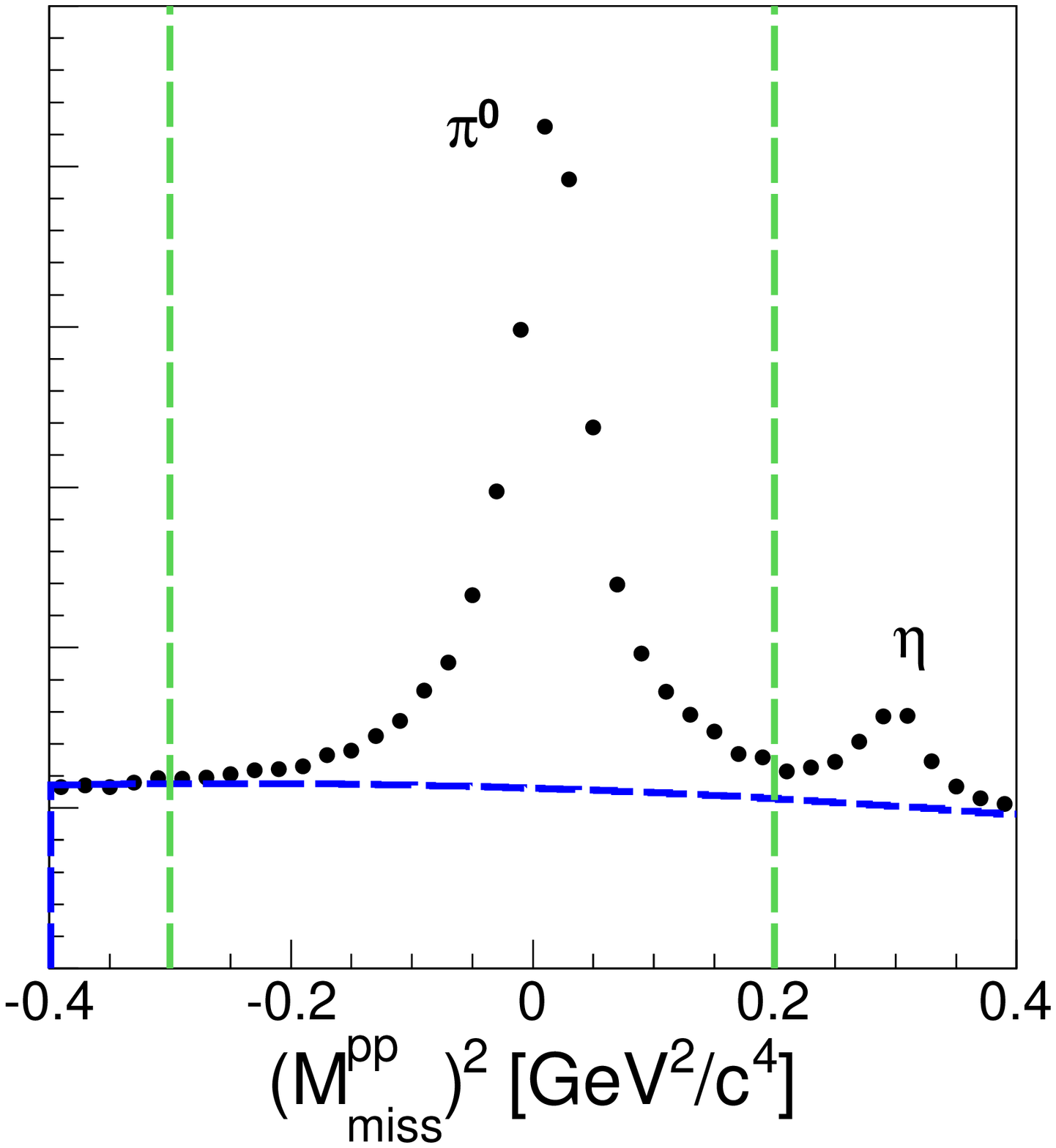}
}

\caption{Left: missing mass squared of the $pp$ system (black dots), simulated two-pion (blue line) and the difference distributions (red points)  after rejection of the elastic proton-proton scattering events. Right: an example of a  fit to the subtracted spectrum in the squared missing mass window (limited by the vertical dashed lines) around the missing mass $\pi^0$ peak.}
\label{ppi0_select}       
\end{figure}

\subsection{Missing mass distributions}

Figure \ref{ppip_select} (left) displays the distribution of missing mass squared of the $p\pi^+$ pair with respect to the beam-target system, where the prominent peak centered around the nominal neutron mass (squared) is clearly visible. In order to extract the yield related to the $p\pi^+n$ final state the background under the peak had to be subtracted. For this purpose a fit function consisting of a polynomial (second and third order were considered) and two Gauss functions accounting for the background and the peak, respectively, were used to fit the experimental distributions. We have checked that such a fit describes the missing mass distributions obtained from simulations (see below) and that the widths of both distributions agree very well. The signal yield was determined as the difference between the measured yield and the fitted background around the missing mass peak. Various background parametrizations and fit ranges were considered to evaluate the systematic error related to the extracted reaction yield. An example of such a fit for the $p\pi^+$ events is presented in Fig. \ref{ppip_select} (right) in the missing mass range used for the signal yield extraction. Typical systematic errors amount to $5-11\%$, depending on the particle momenta and background distributions. The same procedure was applied to determine the signal yield in each bin of various distributions presented below.

Figure \ref{ppi0_select} (left) displays the square of the two-proton missing mass distribution for $2p$ events after rejection of the proton-proton elastic scattering events (see Section 2.4 for details). The background on the right hand side of the $\pi^0$ mass is much higher (black dots) and not well separated from the dominant $\pi^0$ peak. The other two peaks visible on top of the continuum stemming from two-pion production, correspond to the mass squared of $\eta$ and $\omega$ mesons, respectively. The shape of the two-pion contribution (dashed blue line) was obtained from dedicated Monte Carlo simulations (see below), assuming uniform phase space population and with normalization to the measured yield. It was verified that details of the modeling of the two-pion production did not modify the shape of the background and led only to slight changes of its magnitude. In order to extract the signal yield related to the $pp\pi^0$ channel, first the two-pion contribution was subtracted followed by a signal + background fit done in a similar way as in the $p\pi^+$ case. Finally, the yield of the $pp\pi^0$ final state was calculated in the window depicted in Fig. \ref{ppi0_select} (right) as the difference between the measured yield and the fitted background. To correct for a small contribution from the $\eta$, the signal was calculated based on the left half of the $\pi^0$ peak position multiplied by factor $2$. The same procedure was applied to extract the pion production yields as a function of other kinematical variables presented in the next sections.

\begin{figure}
\resizebox{0.5\textwidth}{0.24\textheight}{
  \includegraphics{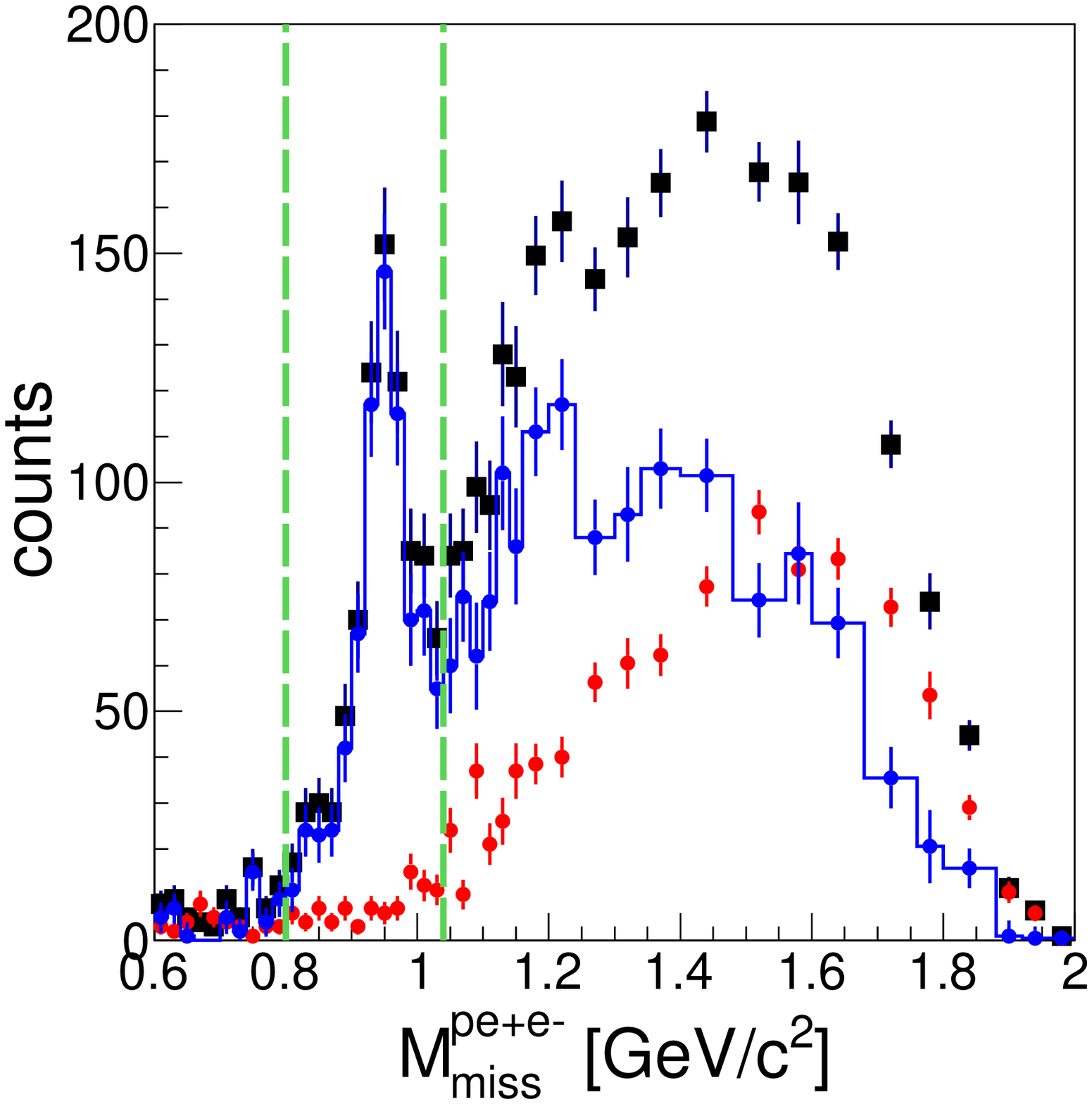}
  \includegraphics{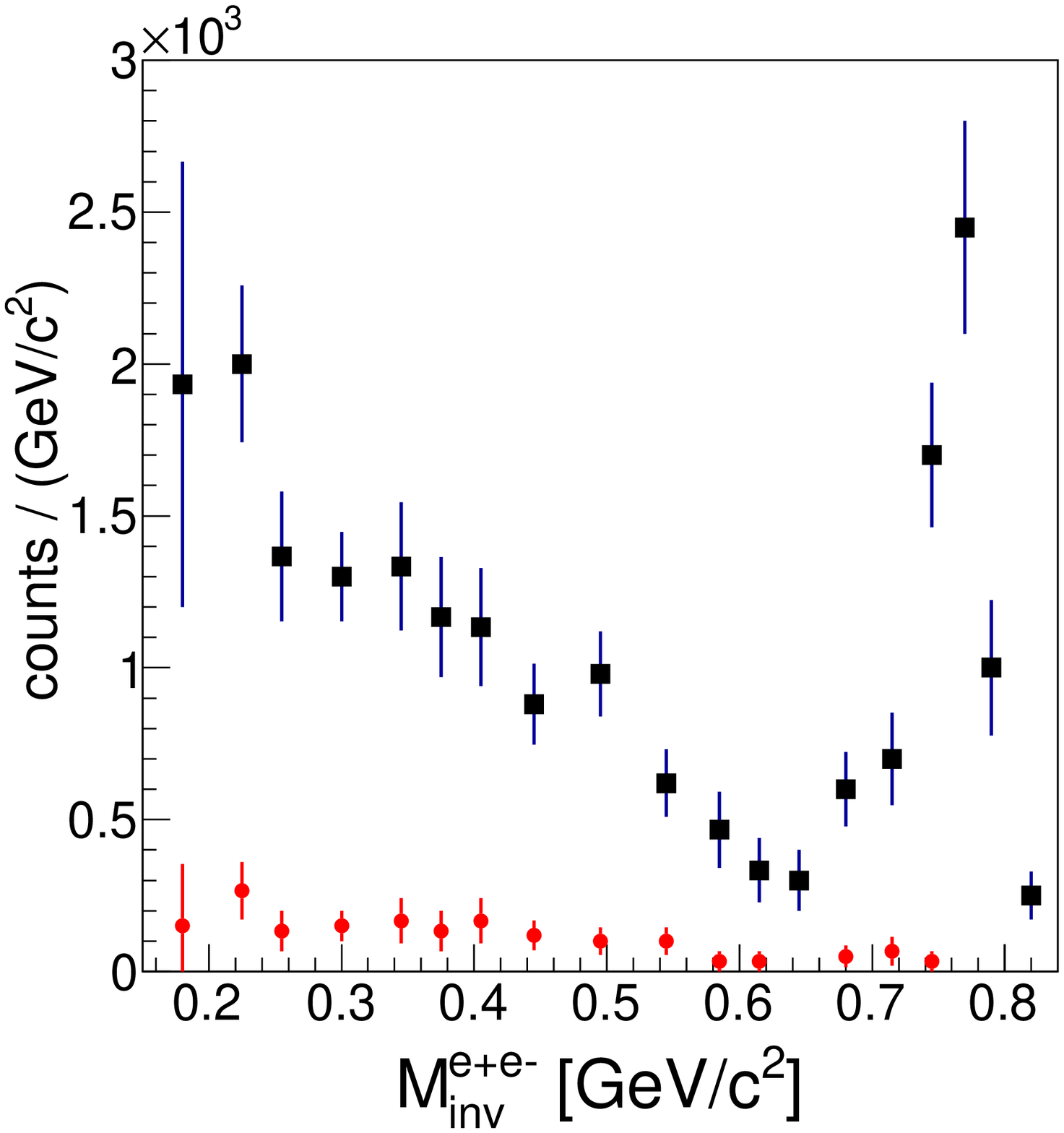}
}
\caption{Left: missing mass distribution for the $pe^+e^-$ system (black squares), sum of $pe^+e^+$ and $pe^-e^-$ (red dots), accounting for the combinatorial background, and the signal $pe^+e^-$ system (blue histogram) for $M_{e^{+}e^{-}} > 0.14$ GeV/$c^{2}$. Right: dielectron invariant mass for the signal pairs (black squares) and the CB (red dots) for the events inside the window around the mass of the missing proton (left panel: limited by the vertical dashed lines, $0.8<M^{pe^+e^-}_{miss}<1.04$ GeV/$c^2$). The total number of signal pairs amounts to 750. Note that the number of counts is given here per $GeV/c^{2}$ to account for the variable bin width used.}
\label{ppee_select}       
\end{figure}

A measurement of any three particles out of four is sufficient for a complete reconstruction of the $ppe^+e^-$ final state. The largest acceptance is achieved for this reaction channel if the detection of one proton and a dielectron is requested. Figure \ref{ppee_select} (left) shows the missing mass distribution of the $pe^+e^-$ system (black squares) together with the CB (a sum of the $pe^-e^-$ and $pe^+e^+$ contributions depicted by red points). The blue histogram presents the signal after the CB subtraction. One should note that the CB contribution increases with the missing mass but it is small in the interesting region around the mass of a missing proton. The right side of Fig. \ref{ppee_select} displays the dielectron invariant mass distributions for events located inside the window centered around the proton mass ($0.8<M^{pe^+e^-}_{miss}<1.04$ GeV/c$^2$) for: (i) the unlike-sign pairs (black squares) and (ii) the CB (red dots) for the $e^+e^-$ pairs with masses $M^{e^+e^-}_{inv} > 0.14$ GeV/c$^{2}$. The latter condition removes abundant pairs originating from the $\pi^0$ Dalitz decay and allows for better inspection of high-mass $e^+e^-$ pairs stemming from the baryon resonance conversions ($R\rightarrow pe^+e^-$) and from vector mesons ($\rho/\omega\rightarrow e^+e^-$) decays. To deduce the yield related to the $ppe^+e^-$ final state and the background contribution, dedicated Monte Carlo simulations, described in the next section, were performed including a realistic detector response and relevant dielectron sources.

\subsection{Normalization}
\label{elastic}

The reaction cross sections were determined using the yield $N_{el}$ of elastic proton-proton scattering measured simultaneously to the other reaction channels. The normalization procedure was described in detail in \cite{pp35GeV}, the overall normalization error was estimated to be $8\%$.

\section{Simulations and acceptance corrections}
\label{sim}

\subsection{Event generation}

Simulations of pion and dielectron production in proton-proton collisions at kinetic energy of $3.5$ GeV were performed by means of the PLUTO event generator \cite{pluto}. A resonance model assuming that the pion production cross section is given by the incoherent sum of various baryon resonance contributions was implemented. We have included all four-star resonances used by Teis et al. \cite{teis} to fit the total one-pion and the $\eta$ meson production cross sections in the range $2.0<\sqrt{s}<5.0$ GeV. As already mentioned, the production amplitudes of the resonances extracted in \cite{teis} are constant and depend neither on the beam energy nor on the resonance production angle, except for the $\Delta(1232)$ resonance for which a strong dependence on the four-momentum transfer from the incoming proton is included in accordance with the OPE results \cite{dmitriev}. So far, the model was however confronted only with data at lower energies \cite{pp2GeV}, where the $\Delta(1232)$ resonance is dominating. We have extended the dependence of resonance production on the production angle to all resonances, as described below. Furthermore, the resonance production cross sections were treated in simulations as free parameters but with fixed isospin relations between production cross sections for the $pn\pi^+$ and the $pp\pi^0$ final states in the respective $I=3/2$ ($\Delta$) and $I=1/2$ ($N^*$) channels (see \cite{pp2GeV}).

Table \ref{resonance_table} summarizes the relevant resonance properties implemented in the simulations: the total decay widths ($\Gamma$), the branching ratios (BR) for $N\pi$ and the $pe^+e^-$ decays (note that the latter ones are defined for the single charge states only). The resonance widths and the $N\pi$ decay branches are adopted from \cite{teis}, except for $N(1535)$, $\Delta(1910)$ and $\Delta(1950)$ the properties of which were taken from \cite{pdg} due to large differences with respect to more recent evaluations. Resonances of similar masses and the same isospin, $I=3/2$ ($\Delta$) or $I=1/2$ ($N^*$), are grouped together in the table for the following reason. In our analysis we identify various resonances by means of the $N\pi$ invariant mass distributions, hence the $\Delta^{++}$ and $N^{*+}$ resonances can be identified as peaks in the $p\pi^+$ and the $n\pi^+$ invariant mass distributions. The resonances grouped together in Table \ref{resonance_table} cannot be isolated by means of the respective $N\pi$ invariant mass distributions because they overlap. In such cases, in the simulations we have selected the resonances (printed in bold style) which have the largest decay branches to the nucleon-pion and to the proton-dielectron final states. In the discussion (see Section \ref{ppee}) of the resulting dielectron yields we have estimated a model uncertainty following from such a selection.

\vspace{0.4cm}
\begin{table}
\begin{tabular}{|ccccc|}
  \hline
  $J^P$ & Resonances & $\Gamma_R$ $[MeV]$ & $BR(N\pi$) & $BR(pe^+e^-)$ \\
  \hline
  \hline
  $3/2^+$ & $\mathbf{\Delta(1232)}$ & 120 & 1 & 4.2e-5 \\
  \hline
  $1/2^+$ & $\mathbf{N(1440)}$ & 350 & 0.65 & 3.06e-6 \\
	\hline
  $3/2^-$ & $\mathbf{N(1520)}$ & 120 & 0.55 & 3.72e-5 \\
	\hline
  $1/2^-$ & $\mathbf{N(1535)}$ & 150 & 0.46 & 1.45e-5 \\
  \hline
  $3/2^+$ & $\Delta(1600)$ & 350 & 0.15 & 0.73e-6 \\
  $1/2^-$ & $\mathbf{\Delta(1620)}$ & 150 & 0.25 & 1.73e-6 \\
  \hline
  $1/2^-$ & $N(1650)$ & 150 & 0.8 & 8.03e-6 \\
  $5/2^-$ & $N(1675)$ & 150 & 0.45 & 1.02e-6 \\
  $5/2^+$ & $\mathbf{N(1680)}$ & 130 & 0.65 & 1.97e-5 \\
  $3/2^+$ & $N(1720)$ & 150 & 0.2 & 3.65e-6 \\
  \hline
  $3/2^-$ & $\mathbf\Delta(1700)$ & 300 & 0.15 & 1.38e-5 \\
  \hline
  $5/2^+$ & $\Delta(1905)$ & 350 & 0.15 & 1.46e-6 \\
  $1/2^+$ & $\mathbf{\Delta(1910)}$ & 280 & 0.25 & 0.73e-5 \\
  $7/2^+$ & $\Delta(1950)$ & 285 & 0.4 & 3.06e-6 \\
  \hline
\end{tabular}
\caption{\small List of resonances and their properties included in the simulations. Some groups of resonances cannot be separated in data. In such a case the resonance with the largest coupling to pion and  dielectron channels (printed in bold) is used in simulations. See the text for details.}
\label{resonance_table}
\end{table}

\vspace{0cm}

For the resonances, the relativistic Breit-Wigner formula with mass dependent widths was used as in \cite{teis}. The branching ratios of the Dalitz decays, given in Table \ref{resonance_table}, are taken from calculations in \cite{zetenyi}, where they are deduced from the known couplings to photons and are defined at the poles of resonances. The full description of the dependency of differential decay widths $d\Gamma_{pe^+e^-}/dm_{e^+e^-}$ on the resonance masses are included in the PLUTO event generator as given by the calculations \cite{zetenyi}. They hold only for a point-like $R N \gamma^*$ coupling and no effects of mass dependent eTFF are included, as for example predicted by VMD models \cite{kriv}. Nevertheless, they can be regarded as a well defined reference to search for effects related to modifications of the resonance-virtual photon vertex due to the intermediate vector meson states.

We have also compared the results of \cite{zetenyi} with other prescriptions for the $\Delta(1232)$ Dalitz decay \cite{kriv,ernst,wolf} used in the dielectron calculations. The disagreement is discussed in \cite{kriv1}. We have found that only the prescriptions of \cite{kriv,zetenyi} consistently reproduce the measured value of the $\Delta(1232)\rightarrow N\gamma$ decay width at $q^2=0$ with the experimentally known magnetic dipole form factor $G_M=3.0\pm 0.05$ \cite{jlab} and electric quadrupole form factor $G_E \approx 0$.

For the angular distributions of the produced resonances we have assumed anisotropic emission in the proton-proton center-of-mass frame depending on the four-mo-mentum transfer$^*$ $t=(p_1-p_R)^{2}$, calculated between the four-momentum vectors of the outgoing resonance ($p_R$) and the incoming nucleon ($p_1$):

\begingroup
\renewcommand{\thefootnote}{\alph{footnote}}
\footnotetext{$^*$ In the calculation of the momentum transfer we have used the following convention for the definition of the incoming proton $p_1$: if the resonance is emitted forward in the CM system, $p_1$ denotes the projectile, otherwise the target proton. }
\endgroup

\begin{equation}\label{ang_d}
d\sigma_R/d t\sim A/t^{\alpha}
\end{equation}where $A$ and $\alpha(M) $ are constants to be derived from the comparison to the data, and $M$ is the respective Breit-Wigner resonance mass. The choice of such a parametrization was motivated by the experimental results on the resonance angular distributions from earlier proton-proton experiments \cite{schlein}, where a strong forward-backward peaking of the resonance production was observed. Moreover, it was found that the anisotropy of the distribution decreases with increasing resonance mass. Such a behavior is expected for peripheral reactions, where the production of heavier resonances requires a larger four momentum transfer and, consequently, a flattening of the angular distributions. The respective $\alpha$ dependency on $M$ has to be, however, found from a comparison to the data.

The decay angular distributions $R\rightarrow N\pi$ of all resonances, except $\Delta(1232)$, have been assumed isotropic, since little is known on the alignment of resonances after production. The $\Delta(1232)$ decay has been modeled proportional to $1+3cos^2(\theta)$, where $\theta$ is the angle of the pion (or nucleon) in the $\Delta$ rest frame with respect to the beam axis. Such a parametrization is predicted by the OPE model and also corroborated by the experimental data \cite{adreev}.

Finally, for the simulation of the dielectron channels, production and decays of the $\eta$, $\rho$ and $\omega$ mesons must be included. The total cross sections of the exclusive $\eta$ and $\omega$ production, $\sigma_{\eta}=140 \pm 14$ $\mu b$ , $\sigma_{\omega}=146 \pm 15$ $\mu b$, respectively, were obtained from a parametrization of the existing data \cite{pp2GeV,tof}. Furthermore, the analysis of the $pp\eta$ Dalitz plot with $\eta$ decaying into $\pi^0\pi^+\pi^-$ from our experiment \cite{khaled_phd} allows for an independent estimate of the $N(1535)$ production. It was found that the contribution of this reaction channel amounts to about $47\%$ and consequently leads to the total production cross section $\sigma_{N(1535)}\simeq 157$ $\mu b$, taking into account the BR($N(1535)\rightarrow N\eta)=0.42$ \cite{pdg}.

The total cross section for $\rho$ meson production was obtained from the $\omega$ cross section by $\sigma_{pp\rho}=0.5\sigma_{pp\omega}$, as observed at $E_{beam}=2.85$ GeV in the DISTO experiment \cite{disto}. This cross section, however, does not account for the off-shell meson production via baryon resonances since it could not be identified in the $\pi^+\pi^-$ invariant mass.

The dielectron decays of the vector mesons were simulated as described in detail in \cite{pp35GeV}. From this work also inclusive cross sections for vector meson production were extracted (at this energy they are larger by a factor 2 than the corresponding exclusive cross sections). They provide important constraints on the total cross sections of the reactions with final states containing additionally one or two pions, for example $pp\pi^0(\pi^0)\omega$, $pn\pi^+(\pi^0)\omega$. They were included in our simulations assuming a production according to phase space distributions.

\subsection{Acceptance and reconstruction efficiency}

To compare the data with the simulation we used a full analysis chain consisting of two steps: (i) processing of the generated events through detectors using the HADES GEANT package and (ii) applying all the reconstruction steps as for the real data \cite{hadesspec}. The normalization of the simulated events was obtained by means of the proton-proton elastic scattering yield which was simulated using the same procedure. The procedure allows for a direct comparison of the measured distributions with the simulated ones within the HADES acceptance. Furthermore, to facilitate fast and easy comparison with the various reaction models, the detector acceptance and the reconstruction efficiencies were calculated and stored in the form of three-dimensional matrices (momentum, polar and azimuthal emission angles) for each particle species ($p$, $\pi^{+}$, $\pi^{-}$, $e^{+}$, $e^{-}$). The acceptance matrices describe the geometrical acceptance of the spectrometer, while the efficiency matrices account for the detection and reconstruction losses within the detector acceptance. The resolution effects were included by means of smearing functions acting on the generated momentum vectors (the matrices and smearing functions are available upon request from the authors). The kinematical cuts related to the channel selections were performed on the filtered events using the same conditions as for the experimental data.

\begin{figure}
\resizebox{0.5\textwidth}{0.22\textheight}{
  \includegraphics{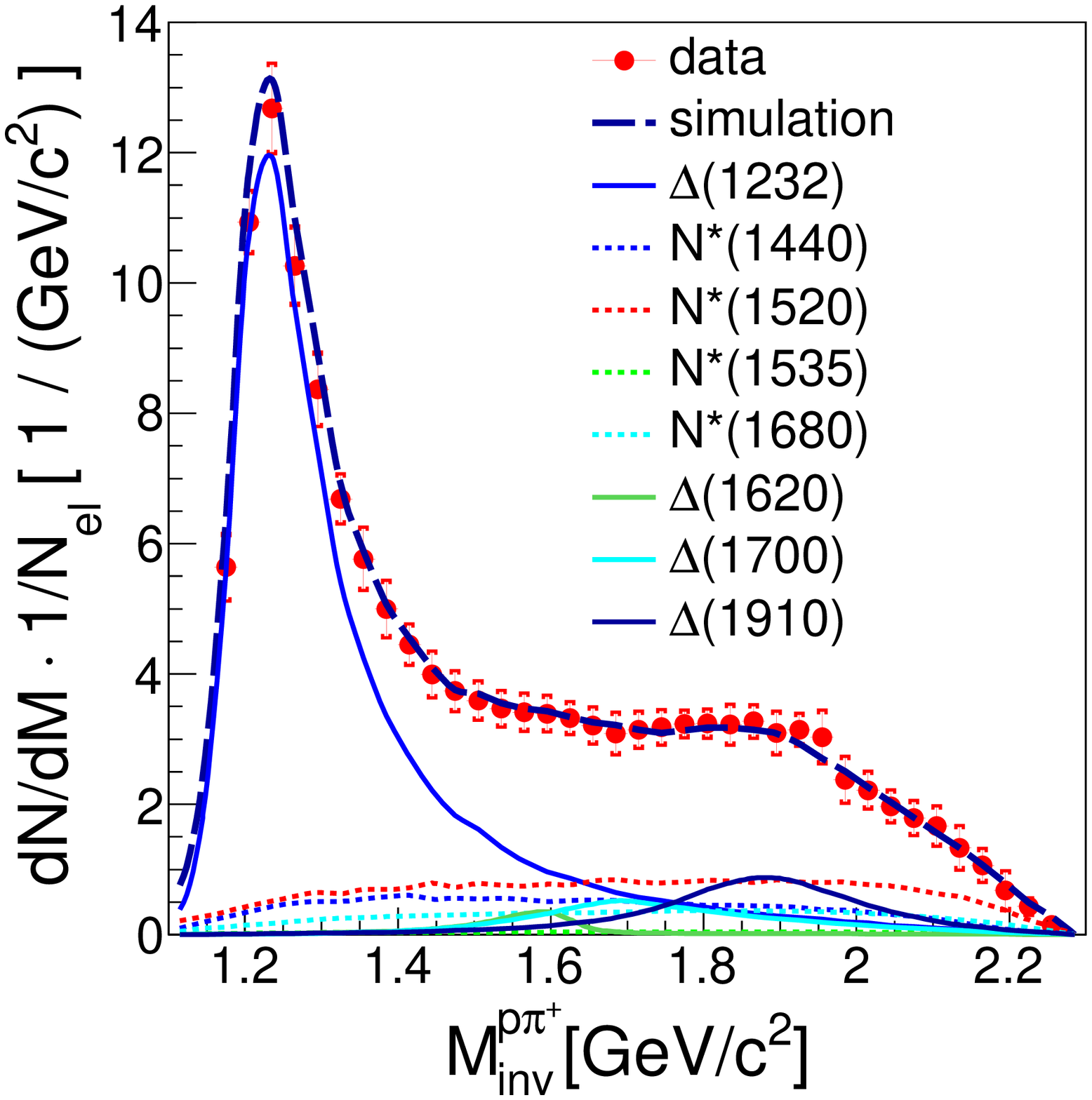}
  \includegraphics{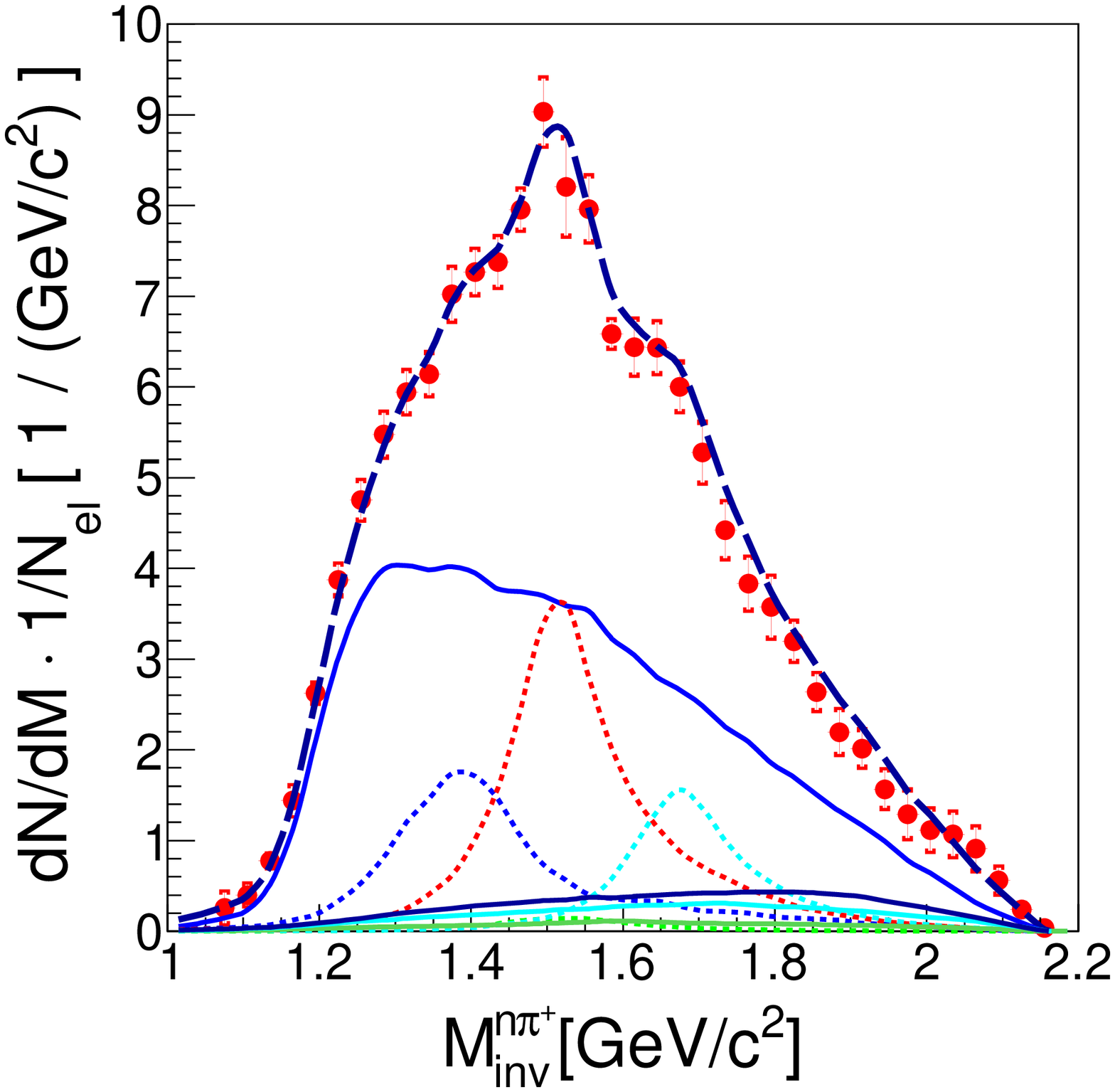}
}
\caption{$np\pi^+$ final state: $p\pi^+$ (left) and $n\pi^+$ (right) invariant mass distributions compared to the result of simulations (dashed curves) assuming an incoherent sum of the resonance contributions shown by separate curves, as indicated in the legend (color code in the online version). The data are normalized to the proton-proton scattering yield $N_{el}$ measured within the HADES acceptance. Indicated error bars are dominated by the systematic errors related to the signal extraction, the constant normalization error ($8\%$) is not included. Normalization to the bin width is applied.}
\label{pnpi_mass}
\end{figure}

In Section \ref{hadrons_hades} we compare various differential distributions for the $pn\pi^+$ and the $pp\pi^0$ final states within the HADES acceptance with the Monte Carlo simulations filtered through the HADES detector by means of the acceptance and efficiency matrices. Since the HADES acceptance is not complete, all acceptance corrections can be performed only by means of a model, which must be proven to be able to describe the data inside the HADES acceptance. Therefore, a detailed comparison of such a model with the data by means of various differential distributions is a mandatory prerequisite for any acceptance corrections and is shown in Section \ref{hadrons_hades} and in the appendix.

\section{$\bf{pn\pi^+}$ and $\bf{pp\pi^0}$ final states}

\label{final states}

\subsection{Distributions within the HADES acceptance}
\label{hadrons_hades}

We start the presentation of our results with the $pp\rightarrow pn\pi^+$ reaction channel. It allows for a separation of the double ($\Delta^{++}$) and the single charged resonances ($\Delta^+,N^{*+}$) by an analysis of the $p\pi^+$ and the $n\pi^+$ invariant mass distributions, respectively. Figure \ref{pnpi_mass} shows the data overlayed with the result of the simulation assuming contributions from the resonances listed in Table \ref{resonance_table}. The data points are normalized to the elastic scattering yields ($N_{el}$) and are displayed together with the errors stemming from the background subtraction procedure, as discussed in Section \ref{channel_selection} (statistic errors are negligible). The normalization error is not included.

\begin{figure*}
\resizebox{1.0\textwidth}{0.25\textheight}{
  \includegraphics{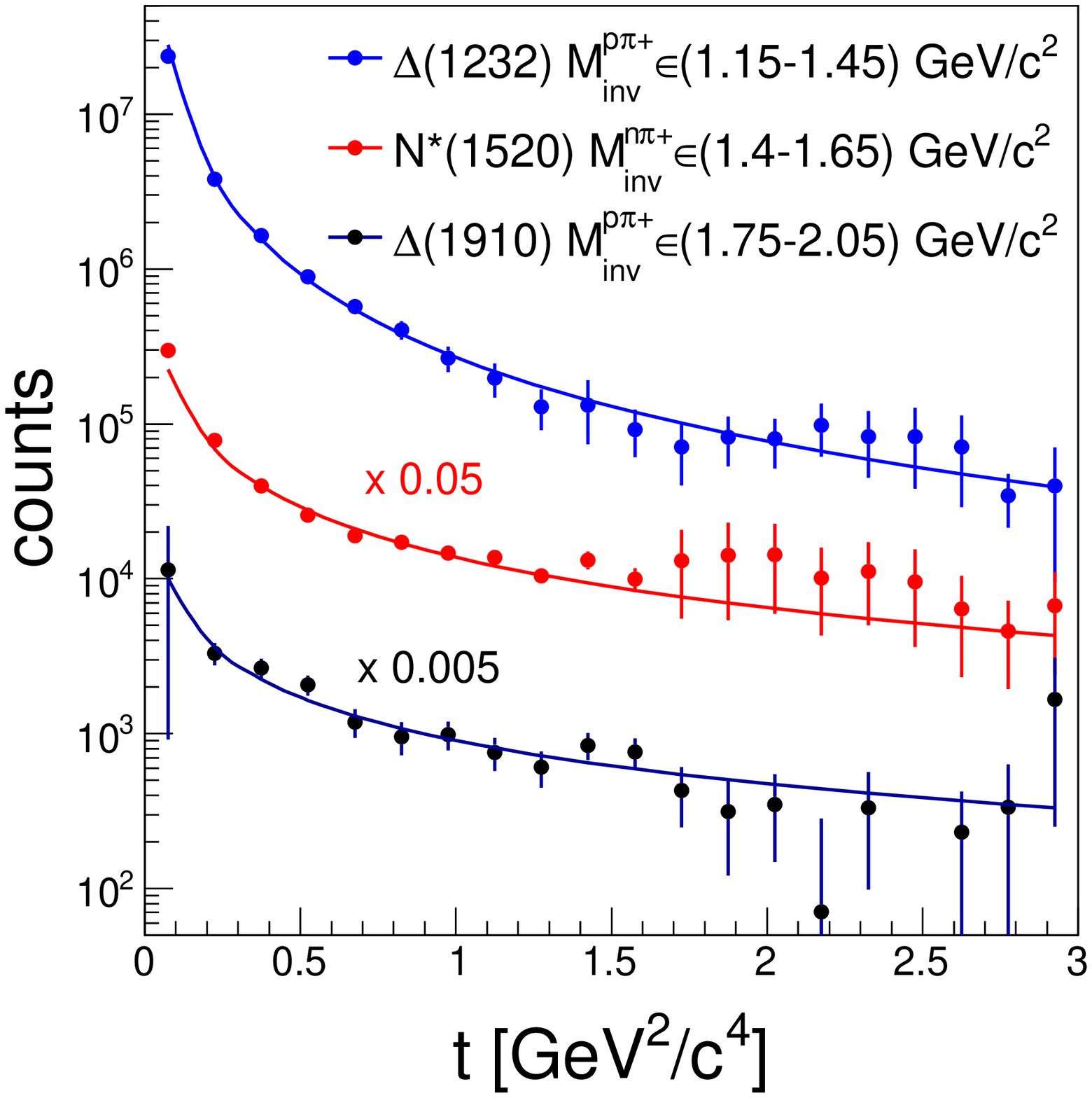}
  \includegraphics{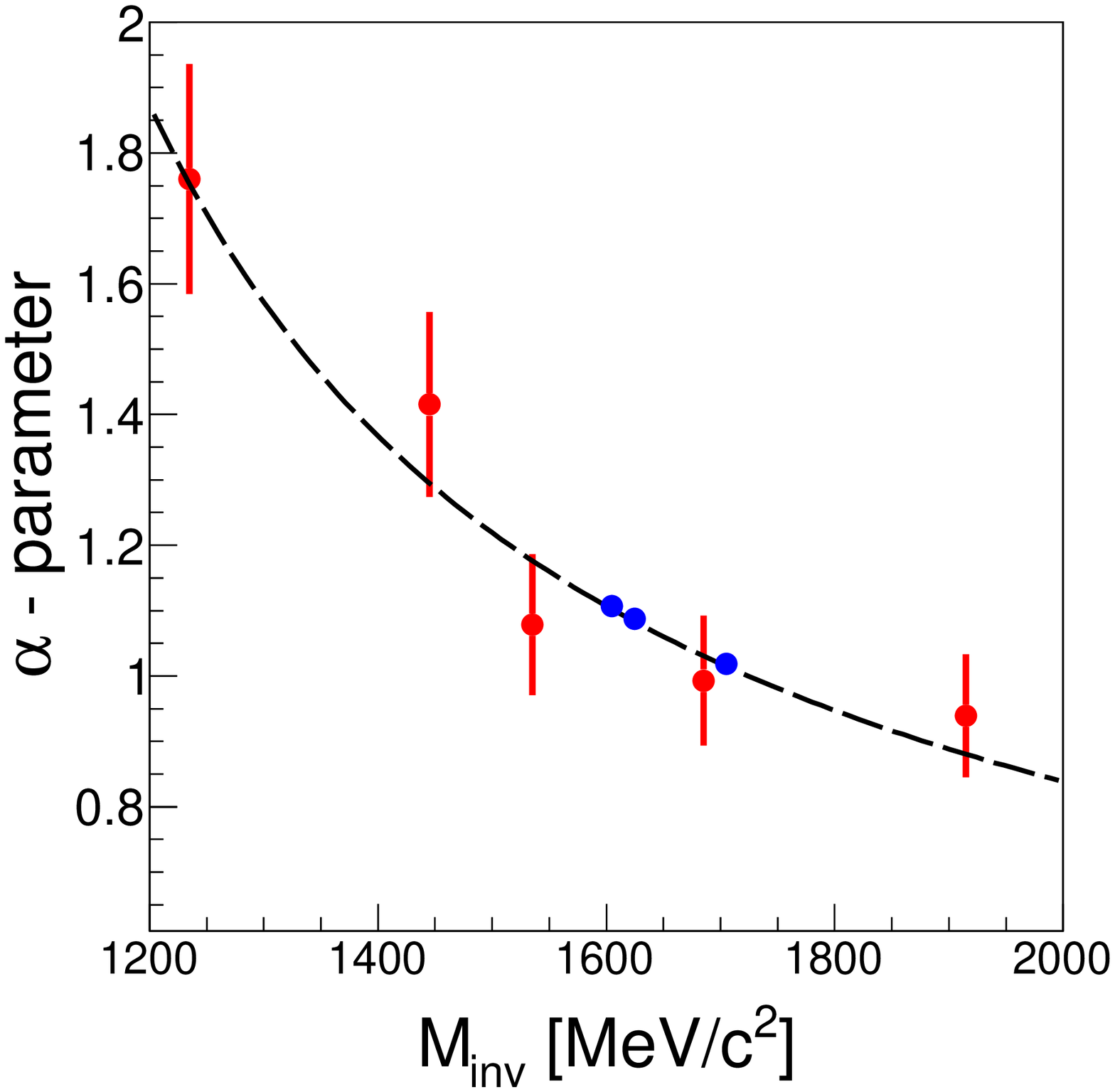}
  \includegraphics{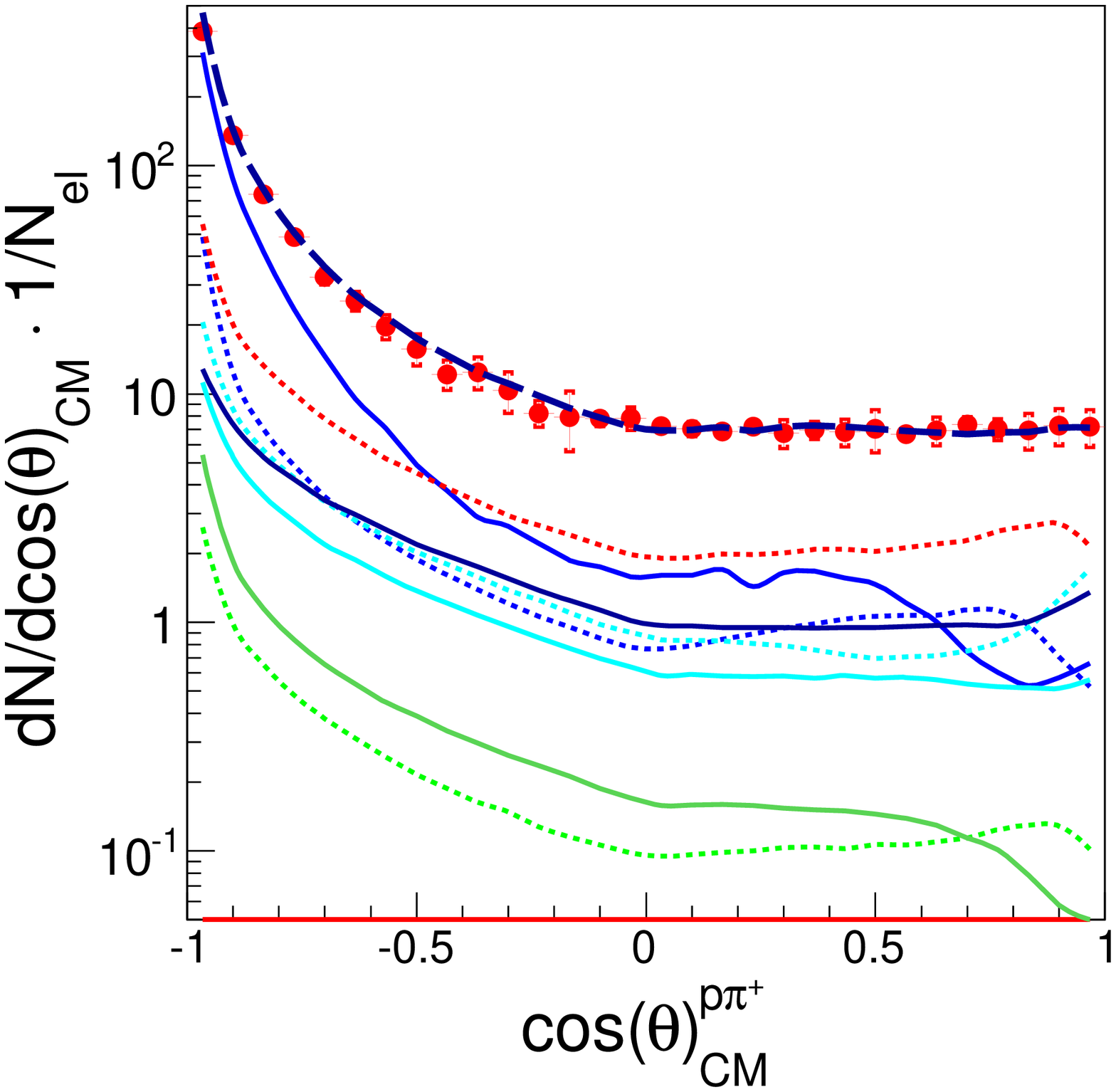}
}
\caption{$np\pi^+$ final state: Left: Acceptance and efficiency corrected distributions of $p\pi^+$ and $n\pi^+$ yield as a function of the four-momentum transfer $t$ compared to fits (solid curves) for the three indicated mass regions. Data from the high mass region are scaled, as indicated, for better visualization. Middle: Dependency of the constant $\alpha$ from Eq.~(\ref {ang_d}) on the resonance mass, obtained from fits to the data (points with errors). The points without errors are the $\alpha$ values deduced from the fit shown by the dashed curve and used in simulations. Right: Center-of-Mass (CM) distribution of the $p\pi^+$ system within the HADES acceptance decomposed into various resonance contributions (same legend as in Fig.~\ref{pnpi_mass}) using the $t$ dependence of the resonance production presented in the middle panel.}
\label{cosppi}       
\end{figure*}

Since the resonance line shapes are fixed in our simulations, the only free parameters, to be found by a comparison to the data, are the resonance production yields and the angular distributions, given by Eq. (\ref{ang_d}). The yields of the resonances were obtained from simultaneous fits to the invariant mass and the four-momentum transfer distributions using an iterative procedure described below. In the first step the $\Delta(1232)^{++}$ resonance, dominating the $p\pi^+$ invariant mass distribution, was considered. In order to extract the slope parameter $\alpha(M)$ for the $\Delta(1232)$, the acceptance and efficiency corrected distribution of the $p\pi^+$ yield as a function of $t$ for the  events with an invariant mass window centered around the resonance pole were plotted, as shown in Fig. \ref{cosppi} (left). The experimental distribution was fitted with a function given by Eq. (\ref{ang_d}) and  the constants   $A(M)$, $\alpha(M)$ were determined. In the next step, the obtained $\Delta(1232)^{++}$ and $\Delta(1232)^{+}$ contributions were subtracted and the same procedure was performed for the $n\pi^+$ events in the region of the $N(1440)$ resonance selected by the respective selection cut on the invariant mass. The yield of the $\Delta^{+}$  was calculated using the isospin relation $\sigma_{\Delta^{++}\rightarrow p\pi^+}=9 \sigma_{\Delta^{+}\rightarrow n\pi^+}$. The sum of both $\Delta$ contributions produces a broad smooth distribution in the $n\pi^+$ invariant mass spectrum, as it can be seen in Fig. \ref{pnpi_mass} (right). On the other hand, the $N^*$ contributions in the $p\pi^+$ invariant mass under the $\Delta(1232)$ peak are very small and influence the fit of the $\Delta^{++}$ angular distribution only marginally.

In a similar manner, the contributions of higher mass resonances $N(1520)^{+}$, $N(1680)^{+}$ and $\Delta(1910)^+$ were extracted in iterative steps. Figure \ref{cosppi} (left) shows the acceptance and efficiency corrected $t$ distributions for the three proton (neutron)-pion mass regions together with the fits and the dependence of the $\alpha$ parameter (middle panel) on the resonance mass extracted from the data. The points with the errors correspond to all investigated resonances, while the points without errors (blue) indicate the values of $\alpha$ deduced from the fit which are used for the other resonances. The observed decrease of $\alpha$ with the resonance mass is equivalent to the flattening of the angular distributions, as also observed in other experiments \cite{schlein}. We have checked that the angular distribution of the $\Delta(1232)$ production obtained from the fit agrees quite well with the one obtained from the already mentioned OPE model of Dimitriev and Sushkov \cite{dmitriev}.

The consistency of the procedure was verified by a simulation with all components included, according to the derived cross sections, given in the next section, and the resonance angular distributions obtained as described above. The acceptance correction of the $t$ distributions has been repeated with the improved model and new $\alpha$ parameters were determined. The second iteration changed only marginally the fit parameters. The final decomposition (here within the HADES acceptance) of the simulated $p\pi^+$ yield as a function of $cos(\theta_{CM}^{p\pi^+})$ into individual contributions from the resonances is displayed in Fig. \ref{cosppi} (right). The asymmetric shape of the angular distribution is due to the acceptance favoring the detection of $p\pi^+$ pairs emitted in the CM in backward direction (or, equivalently, $n\pi^+$ pairs in forward direction). The HADES acceptance and reconstruction efficiency increase as a function of the resonance mass from $6\%$ to $15\%$.

\begin{figure}
\resizebox{0.24\textwidth}{0.22\textheight}{
  \includegraphics{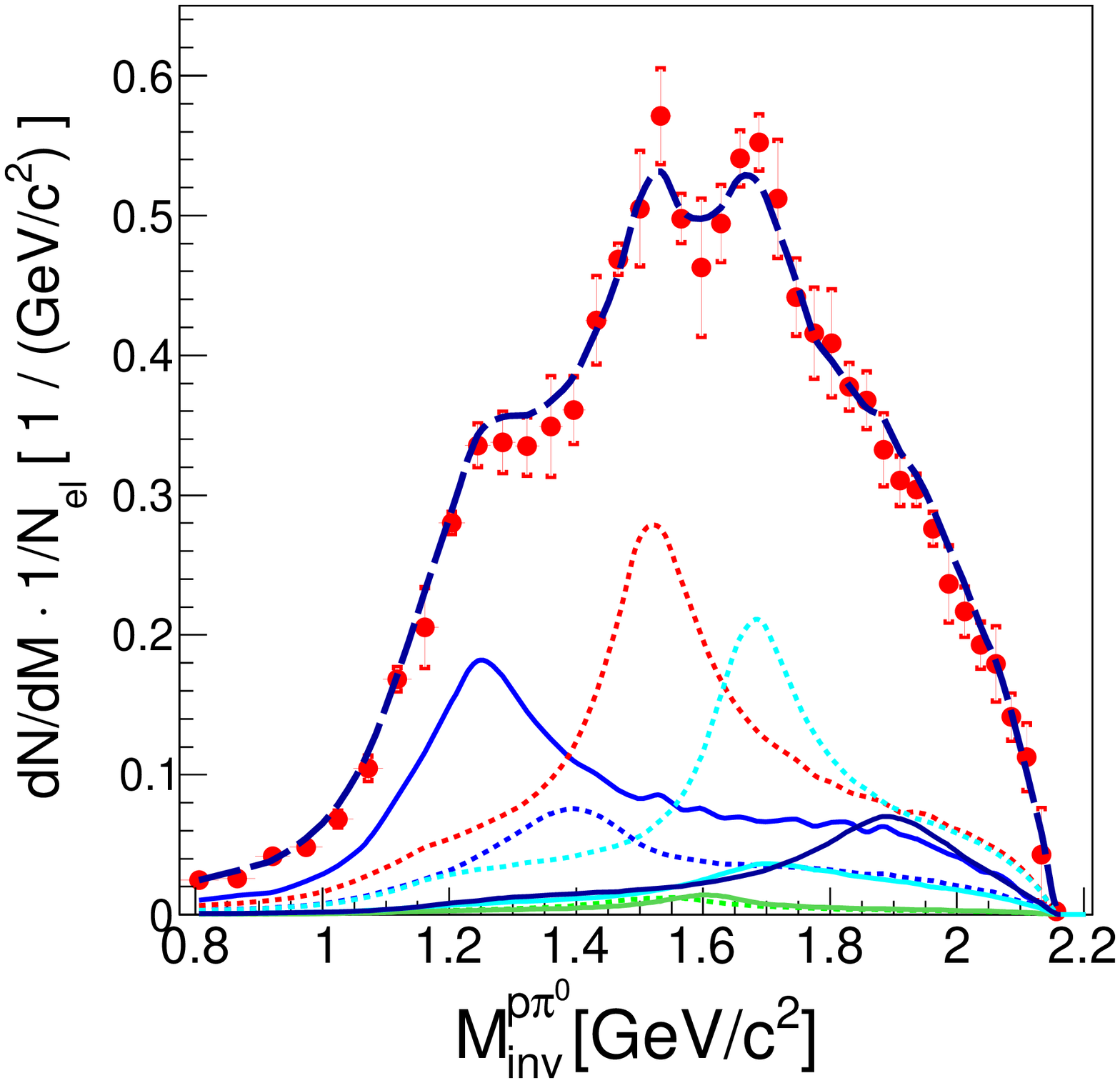}
  }
\resizebox{0.24\textwidth}{0.22\textheight}{
  \includegraphics{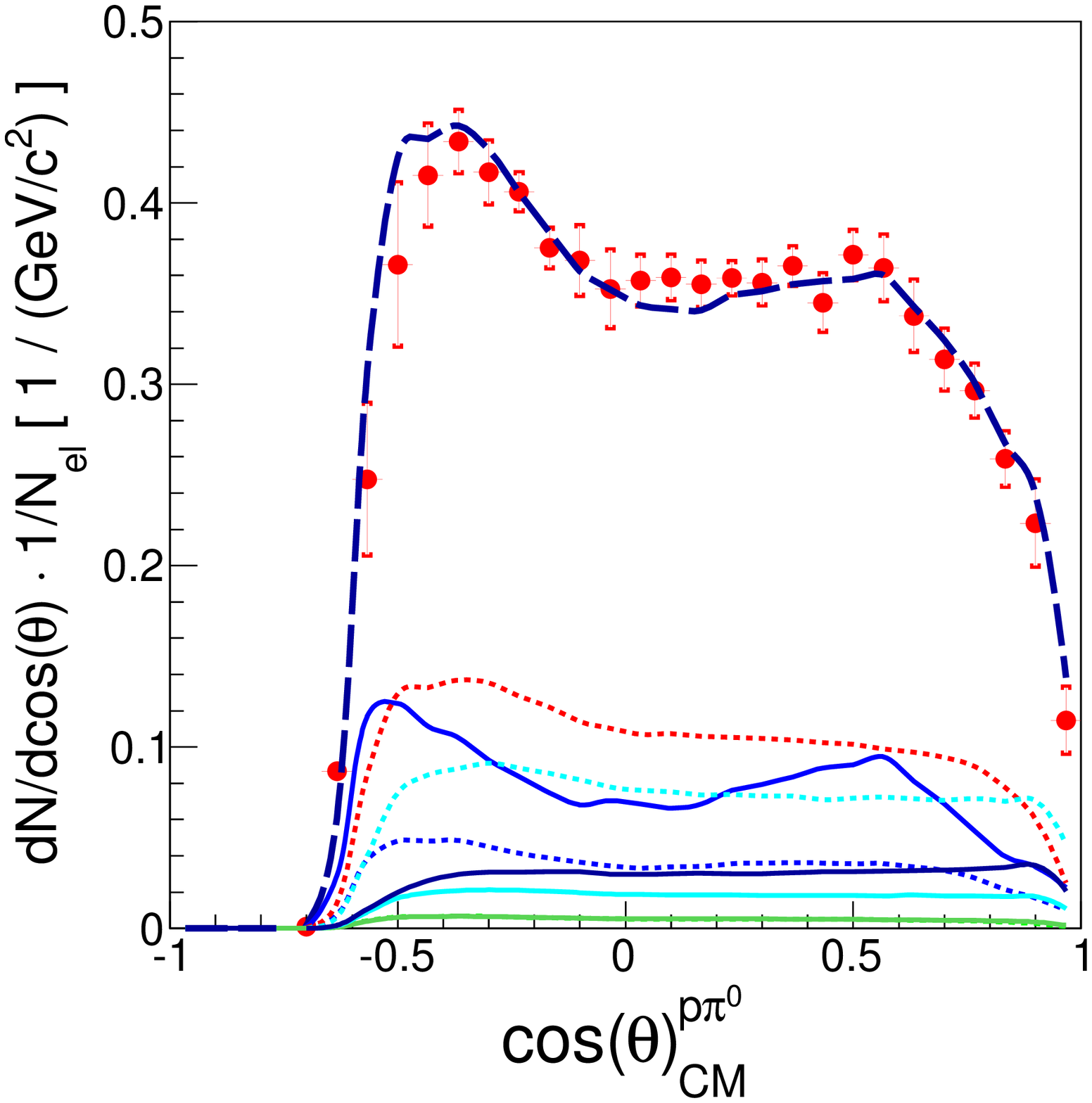}
}
\caption{$pp\pi^0$ final state: $p\pi^0$ invariant mass distribution (left) and the CM angular distributions (right) compared to the result of the simulation (line style as in  Fig.~\ref{pnpi_mass}, normalization to the bin width is applied).}
\label{pppi0_mass}       
\end{figure}

Finally, the extracted resonance yields and the angular distributions were included in the simulation of the $pp\rightarrow pp\pi^0$ reaction channel. In our model, the cross sections for the $pn\pi^+$ and $pp\pi^0$ final states are fixed by their isospin relations, hence no additional scaling is allowed. Indeed, a very good agreement between simulation and the data was also achieved for this reaction channel. Figure~\ref{pppi0_mass} presents a comparison of the $p\pi^0$ invariant mass and the CM angle distributions of the $p\pi^0$ system, obtained in the experiment, with the results of the simulation.  Since the two final-state protons are undistinguishable, both combinations of protons with a neutral pion were included in the presented distributions by taking two possible combinations per event (each with a weight 0.5). Contrary to the $pn\pi^+$ final state, the intensity of the $\Delta(1232)$ resonance is reduced and the contributions of higher mass resonances are more pronounced. One should note, however, that the distributions are strongly affected by the HADES acceptance which is in general smaller by a factor $2-3$, depending on the $p\pi^0$ mass, as compared to the acceptance for the $pn\pi^+$ final state. In the angular distributions for the two reaction channels (right panels of Figs.~\ref{cosppi} and \ref{pppi0_mass}), a clear cut-off is visible in the $p\pi^0$ case. While the acceptance for the $pn\pi^+$ channel is large for the backward emitted $p\pi^+$ pairs the acceptance for the $pp\pi^0$ is strongly reduced in this region. Consequently, $p\pi^0$ events from reactions characterized by small momentum transfer are suppressed with respect to the $p\pi^+$ case.

To perform more detailed comparisons between the data and the model we have also investigated angular distributions defined in the Gottfried-Jackson ($GJ$) and the helicity ($H$) reference frames. The respective distributions are presented in the Appendix and show overall good agreement with our model. The distributions in the $GJ$ reference frame are related to the decay angles in the resonance rest frame which in particular corroborate our assumptions about the $\Delta(1232)$ decay (see Figs. \ref{GJH_pnpi1} and \ref{GJH_ppi}).

\subsection{Acceptance corrected cross sections}
\label{hadrons_corrected}

Based on the studies presented in the previous section, we conclude that our simulation reproduces the data satisfactorily. Therefore the simulation can be used to correct the data for losses due to limited acceptance and inefficiencies of the detection and the reconstruction processes. Acceptance corrected distributions can then be compared to other reaction models than those used in the simulation. The correction factors were calculated from the simulations as the ratio between the generated and the accepted and reconstructed distributions as one dimensional functions for all studied kinematical variables separately (i.e the invariant masses and the various angles discussed in the previous section). In this chapter we present only some selected distributions.

\begin{figure}
\resizebox{0.5\textwidth}{0.22\textheight}{
  \includegraphics{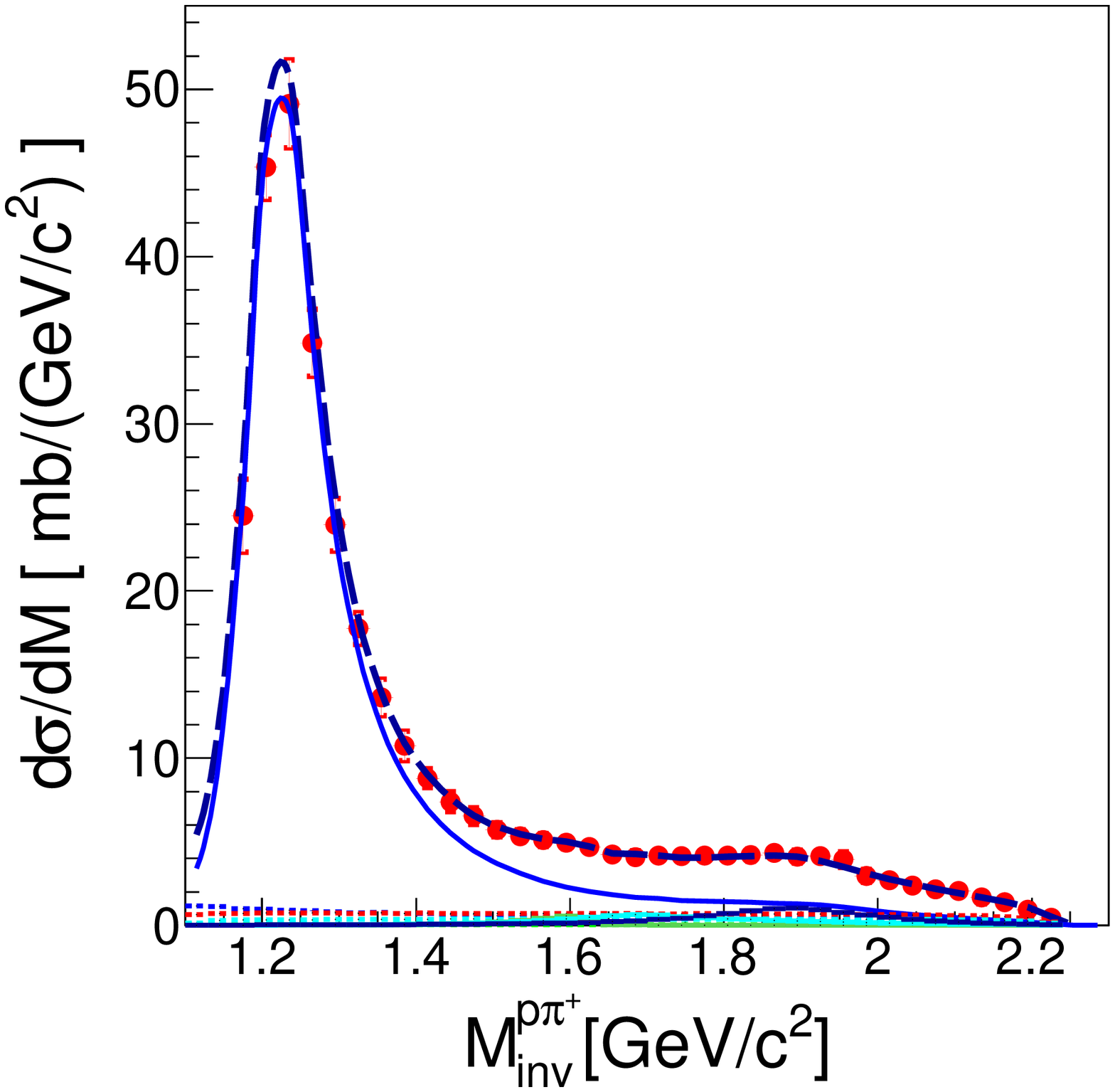}
  \includegraphics{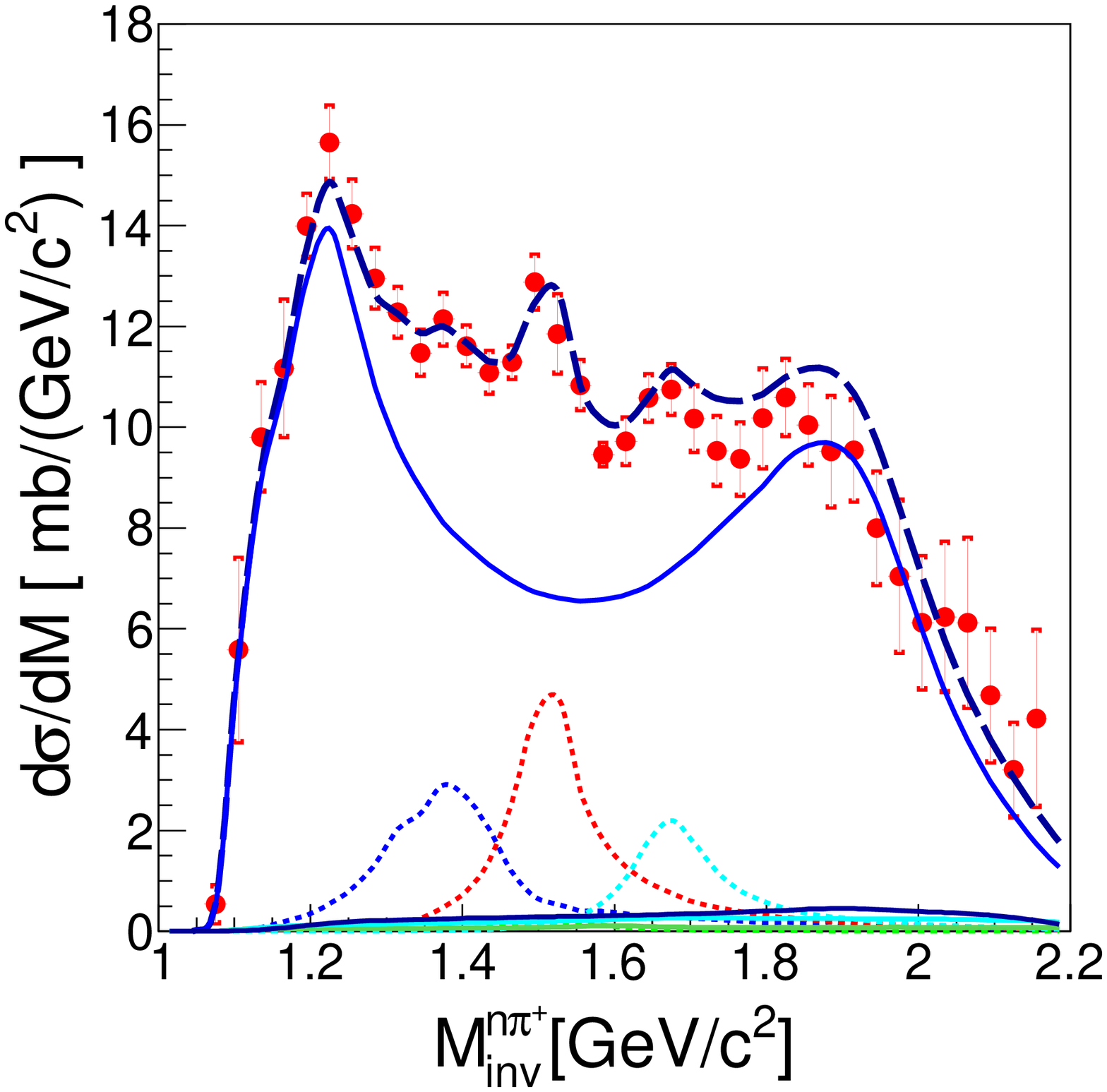}
}
\caption{$np\pi^+$ final state: Acceptance corrected $p\pi^+$ (left) and $n\pi^+$ (right) invariant mass distributions compared to the simulation result (dashed curves). Resonance contributions are shown separately (line style as in Fig.~\ref{pnpi_mass}).}
\label{pnpi_mass_corr}       
\end{figure}

Figure~\ref{pnpi_mass_corr} displays the acceptance and efficiency corrected charged pion differential cross sections as a function of the $p\pi^+$ and the $n\pi^+$ invariant masses for the $pn\pi^+$ final state. The distributions are overlayed with the simulation decomposed into contributions of the $\Delta$ and the $N^*$ resonances, indicated as in the previous Figs.~\ref{cosppi}-\ref{pppi0_mass}. One can notice, by comparing to the respective uncorrected distributions shown in Fig.~\ref{pnpi_mass}, that the corrections enhance the low-mass $\Delta(1232)$ region for the $p\pi^+$ and $n\pi^+$ systems and the high-mass region ($M_{n\pi^+}>1.9~GeV/c^2$) for the $n\pi^+$ system. The salient feature of the $p\pi^+$ system is, as already observed in the uncorrected spectra, a dominant $\Delta(1232)^{++}$ contribution and a slight enhancement around $M_{p\pi^+}=1.9$ GeV/c$^2$ which may indicate contributions from the higher mass $\Delta$ states. The line shape of the $\Delta(1232)^{++}$, which dominates the $p\pi^+$ invariant mass distribution up to $1.6$ GeV/c$^2$, is perfectly described by our simulation. This observation is important in view of the various parameterizations of the resonance spectral function used in transport models which substantially differ at high $\Delta$ masses as discussed in \cite{elena}. Our fit supports a parametrization of the total width based on the Moniz model \cite{moniz} which strongly suppresses the high-mass tail of the resonance (see \cite{elena} for details).

The $n\pi^+$ invariant mass distribution reveals also contributions of the single-charged resonances: $\Delta(1232)^+$, $N(1440)$, $N(1520)$ and $N(1680)$. This region is, however, dominated by $n\pi^+$ pairs from the $\Delta(1232)^{++}n \rightarrow p\pi^+n$ final state and is characterized by a continuous invariant mass distribution with an enhancement around $1.9$ GeV/c$^2$.  It is interesting to note that the enhancement is due to the assumed anisotropy of the $\Delta(1232)^{++}$ decay $1+3cos^2(\theta)$ which is also corroborated by the angular distributions obtained in the $GJ$ frame (see Fig. \ref{GJH_pnpi1} in the Appendix). Note that the $\Delta(1232)$ contribution shown in Fig.~\ref{pnpi_mass_corr} presents the sum of $\Delta(1232)^{++}$ and $\Delta(1232)^+$, where the latter resonance peaks approximately at the pole position. It is particularly important to note the strong contributions of the $N(1520)$ and $N(1680)$ resonances which are relevant for dielectron production because of their relatively large Dalitz decay branching ratios (see Table~\ref{resonance_table}).

\begin{figure}
\resizebox{0.5\textwidth}{0.22\textheight}{
  \includegraphics{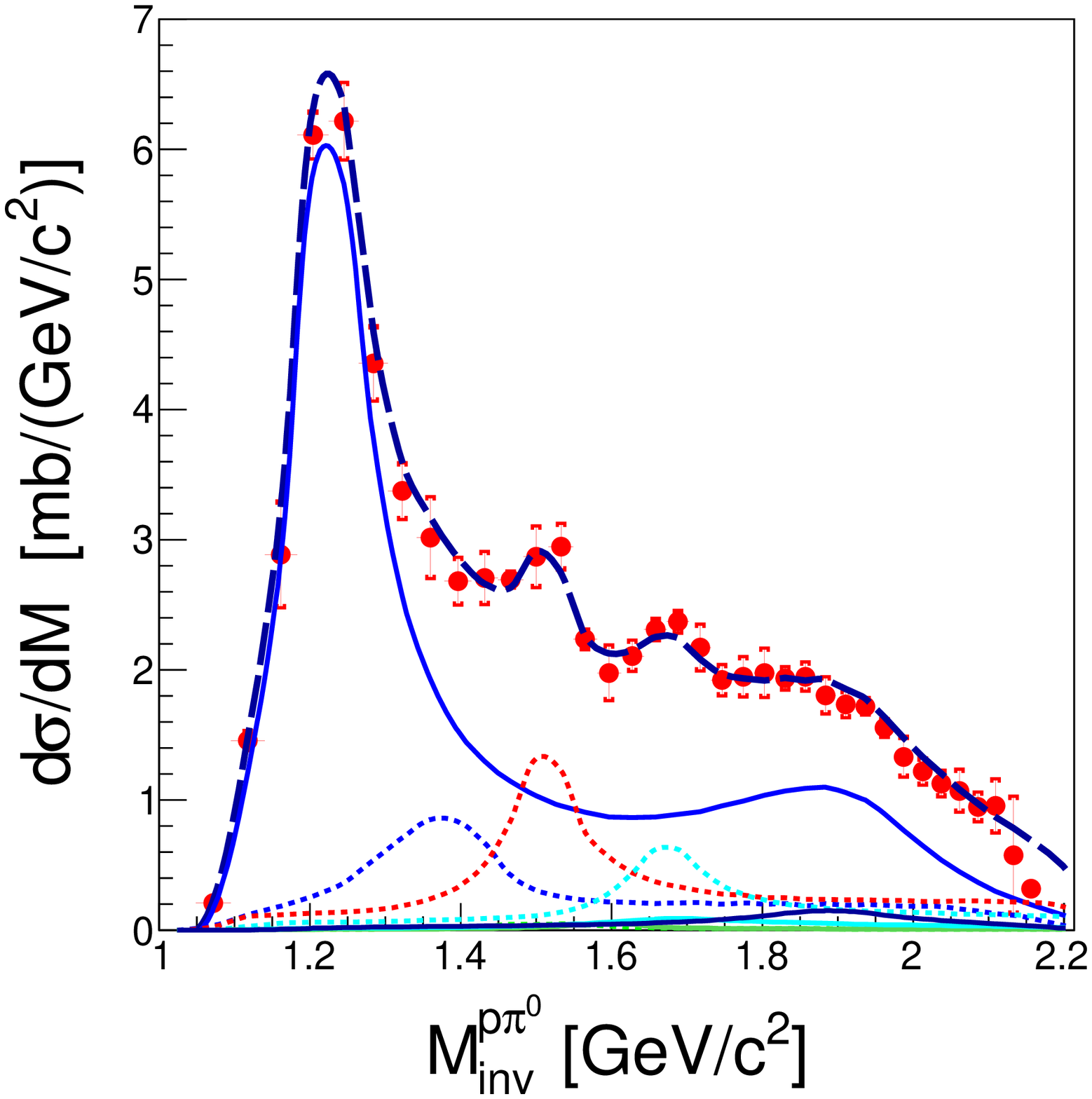}
  \includegraphics{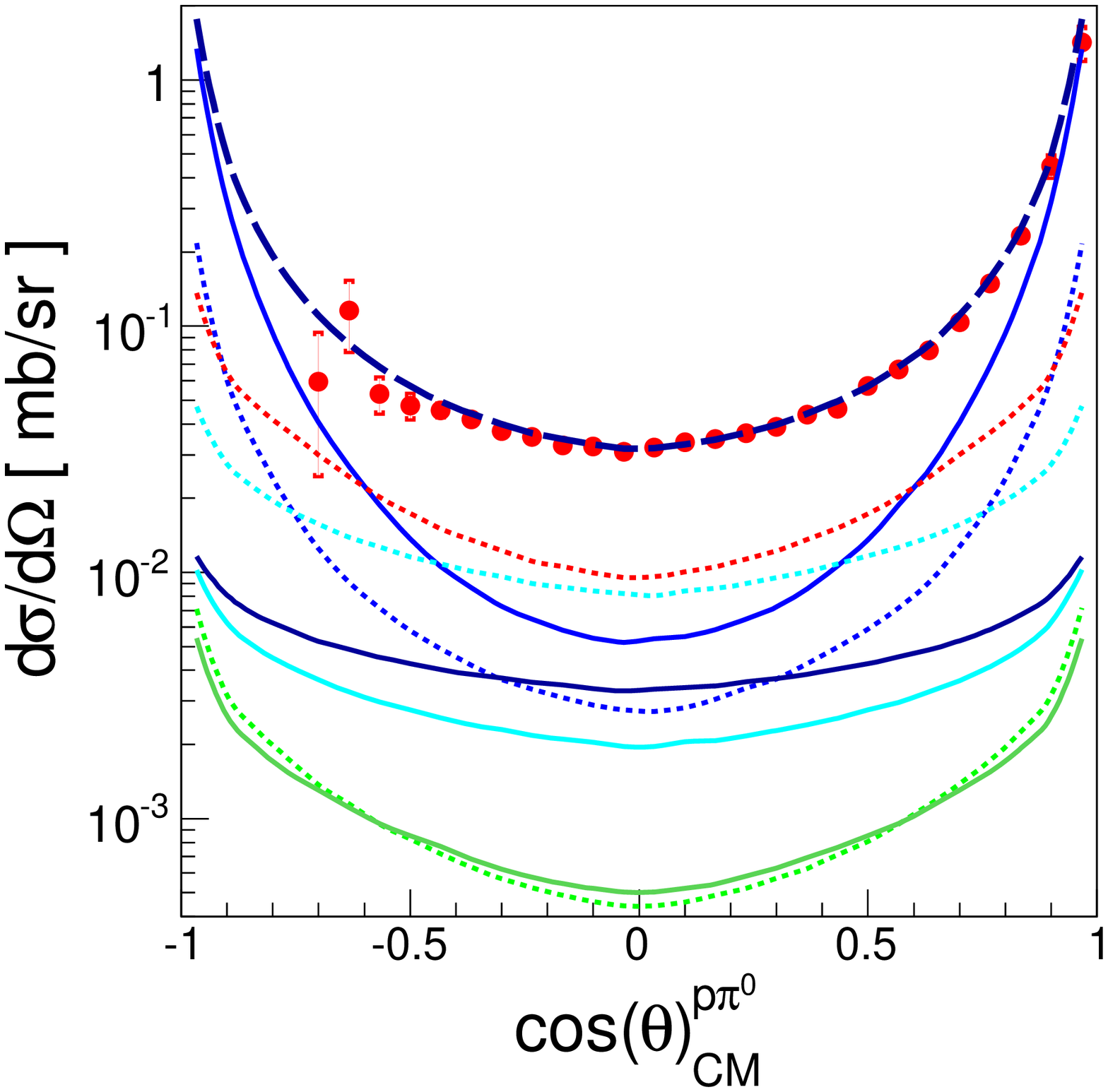}
}
\caption{$pp\pi^0$ final state: Acceptance corrected $p\pi^0$ invariant mass (left) and the CM angular distributions (right)  compared to the simulation result (dashed curves). Resonance contributions are shown separately (line style as in Fig.~\ref{pnpi_mass}).}
\label{pppi0_mass_corr}       
\end{figure}

The acceptance corrected invariant mass distributions for $pp\pi^0$  final states are shown in Fig.~\ref{pppi0_mass_corr} together with the simulation results. In contrast to the $pn\pi^+$ reaction channel, the $pp\pi^0$ final state is sensitive only to the contributions of single-charged resonances, hence the very strong signal from the  double-charged $\Delta(1232)^{++}$ is absent and other resonances are more prominent. On the other hand, a disadvantage of this channel is that the final state of two protons does not allow for a unique reconstruction of the resonance mass and leads to a slight spectral distortion due to averaging between two possible pion-proton combinations. Nevertheless, the enhancements around $N(1520)$ and $N(1680)$ are also clearly visible, as it is the case in the $pp\rightarrow pn\pi^+$ reaction channel. Figure \ref{pppi0_mass_corr} (right) shows the differential cross section as a function of the CM angle of the proton-pion system in comparison to our model calculations. The expected strong anisotropy, decreasing with increasing resonance mass of the $p\pi^0$ production, is clearly visible (see the components). The lack of data points below $cos(\theta^{p\pi^0}_{CM}) < -0.6$ reflects the acceptance losses in the HADES spectrometer.

From the acceptance corrected spectra the total cross sections for the $pp\pi^0$ and the $pn\pi^+$ final states can be calculated. They have been obtained as an average of the integrated differential cross sections expressed as a function of the pion-nucleon invariant mass and various angles presented above. The respective cross sections amount to $\sigma_{pp\pi^0}=2.50\pm 0.23 (syst) \pm 0.2 (norm)$ mb and $\sigma_{pn\pi^+}=10.69\pm 1.2 (syst) \pm 0.85 (norm)$ mb (the statistical errors are negligible). The systematic errors were estimated from the differences between the integrated differential cross sections obtained after the respective acceptance corrections on the above mentioned distributions.

The $pp\pi^0$ distributions presented above are particularly interesting since they provide a direct input to calculations of the resonance conversion $R\rightarrow pe^+e^-$. However, as discussed above, in our simulation we have used only a subset of resonances because we cannot distinguish between overlapping states in the pion-nucleon invariant mass distributions. Nevertheless, using the resonance model ansatz we are able to extract upper limits on contributions from other possible resonances within the given groups in Table \ref{resonance_table} and can calculate the respective uncertainty of the dielectron yield. For this purpose we have repeated our simulations substituting the selected resonance with other resonances, one by one, belonging to the same group (see Table \ref{resonance_table}) but keeping the other components in the simulations unchanged. The obtained cross sections are listed in the second column of Table \ref{resonance_results}. The error in the determination of the cross section for production of resonances were estimated for each resonance separately from the pion-nucleon invariant mass distributions by changing  the respective yield within the experimental error bars but with all other components fixed. The relative errors for some resonances are quite large due to their small contribution to the pion production, leading to a limited sensitivity.

\vspace{0.4cm}

\begin{table}
\begin{tabular}{|cccc|}
  \hline
  Resonances & $\sigma_R$ & $\sigma_R^{Teis} (\sigma_R^{GiBUU})$ & $\sigma_R^{UrQMD}$ \\
  \hline
  \hline
  $\mathbf{\Delta(1232)}$ & $2.53 \pm 0.31 $ & 2.0 (2.2) & 1.7\\
  \hline
  $\mathbf{N(1440)}$ &  $1.50 \pm 0.37 $  & 0.83 (3.63) & 1.15\\
	\hline
  $\mathbf{N(1520)}$ &  $1.8 \pm 0.3 $  & 0.22 (0.27) &  1.7 \\
	\hline
  $\mathbf{N(1535)}$ &  $0.152 \pm 0.015$  & 0.53 (0.53) & 0.8\\
  \hline
  $\Delta(1600)$          &  $< 0.24 \pm 0.10$ & 0.70 (0.14) & 0.4 \\
  $\mathbf{\Delta(1620)}$ &  $< 0.10 \pm 0.03$ & 0.60 (0.10) & 0.2 \\
  \hline
  $N(1650)$               & $< 0.81 \pm 0.13$ & 0.23 (0.24) & 0.4\\
  $N(1675)$               & $< 1.65 \pm 0.27$ & 2.26  (0.94) & 1.2 \\
  $\mathbf{N(1680)}$      & $< 0.90 \pm 0.15$  & 0.21 (0.22)  & 1.2 \\
  $N(1720)$               & $< 4.41 \pm 0.72$ & 0.15  (0.14) & 0.68 \\
  \hline
  $\mathbf{\Delta(1700)}$   & $0.45 \pm 0.16$  & 0.10 (0.06) & 0.35\\
  \hline
  $\Delta(1905)$            & $< 0.85 \pm 0.53$ & 0.10 (0.06) & 0.25\\
  $\mathbf{\Delta(1910)}$   & $< 0.38 \pm 0.11$ & 0.71 (0.14) & 0.08 \\
  $\Delta(1950)$            & $< 0.10 \pm 0.06$  & 0.08 (0.10)  & 0.25 \\
  \hline
\end{tabular}
\caption{\small Cross sections in units of $mb$ for the single positively charged resonances extracted from our data (second column), the Teis et al. model \cite{teis} (third column) and used in the GiBUU \cite{GiBUU} (number in brackets in the third column) or the UrQMD \cite{urqmd} (fourth column).}
\label{resonance_results}
\end{table}

\begin{figure}
\resizebox{0.5\textwidth}{0.22\textheight}{
  \includegraphics{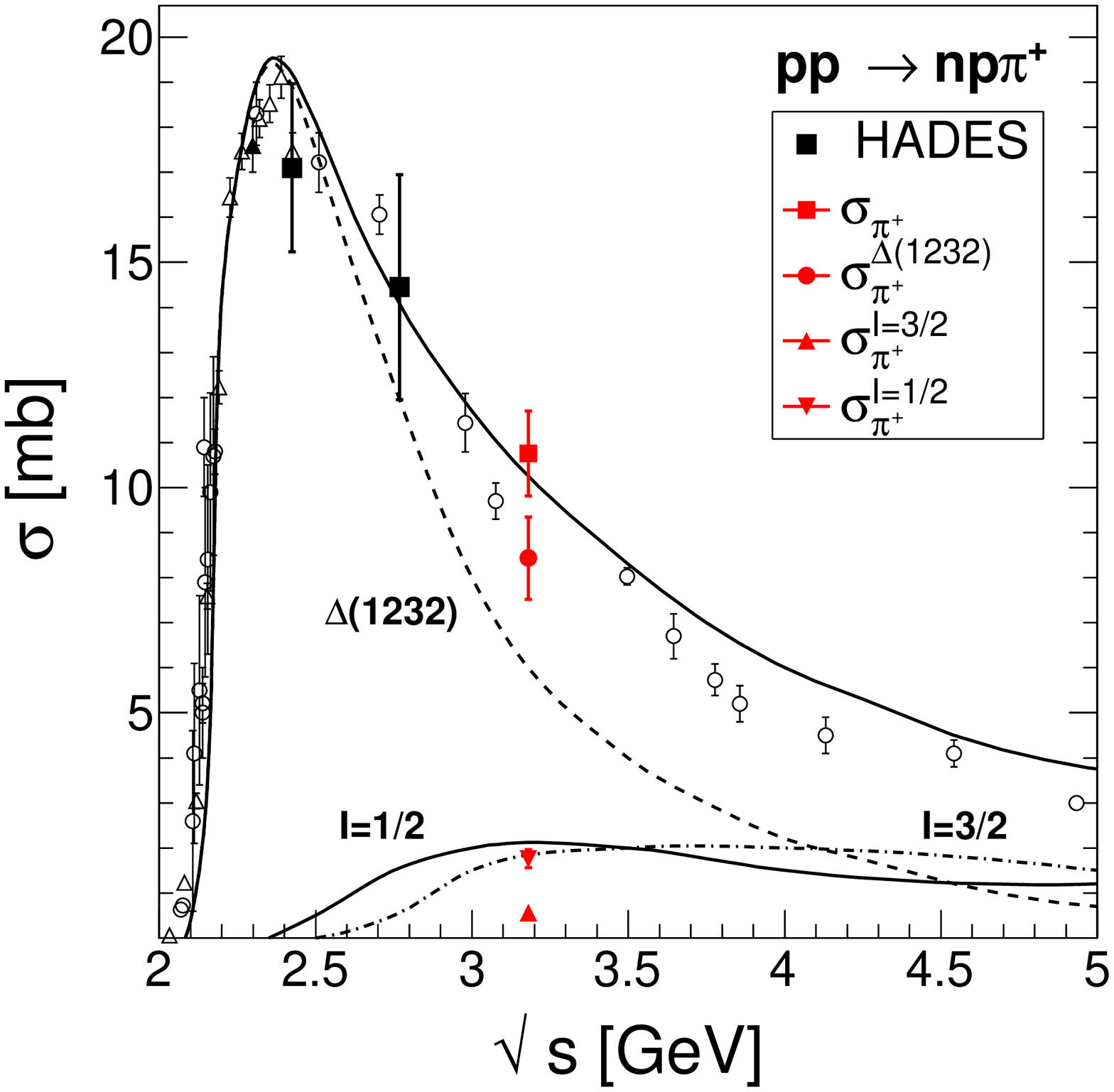}
  \includegraphics{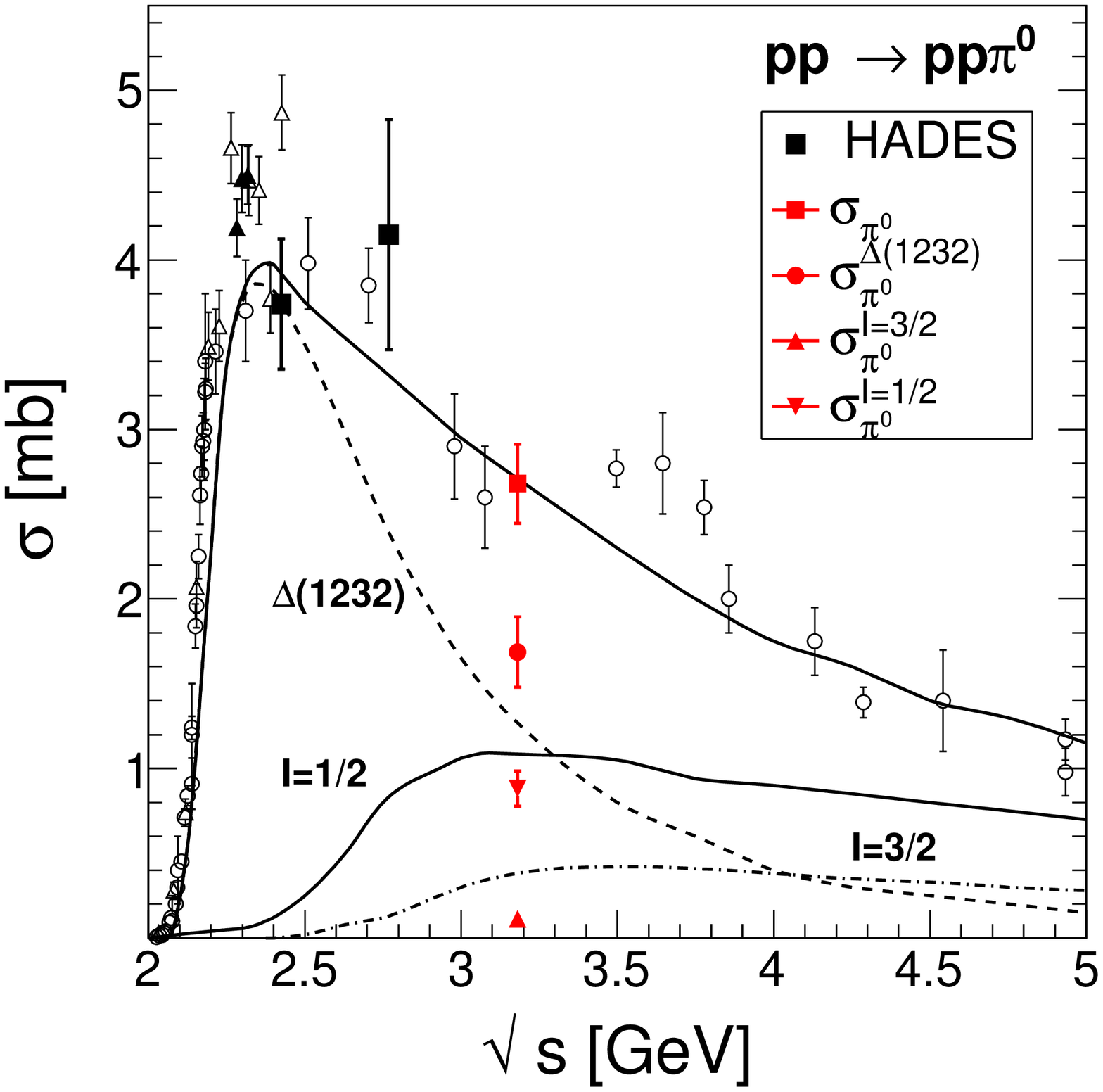}
}
\caption{One-pion (left: charged, right: neutral) exclusive cross sections as a function of the total CM energy $\sqrt{s}$ separated into contributions of the $\Delta(1232)$, the higher mass $I=3/2$ ($\Delta$) and the $I=1/2$ ($N^*$) resonances in comparison to the parametrization \cite{teis} and other experimental data. The data compilation is taken from \cite{teis}. The HADES results at $\sqrt{s}$ = 3.18~GeV are depicted as full symbols (black squares from the measuremets at lower energies \cite{pp2GeV}).}
\label{pnpi_xs}       
\end{figure}

The last two columns in Table \ref{resonance_results} present the resonance cross sections from the model of \cite{teis} and the modified values used in the GiBUU code \cite{GiBUU} (values in brackets), as well as the values used in the UrQMD \cite{urqmd} code. Figure~\ref{pnpi_xs} shows the total one-pion exclusive cross sections as a function of $\sqrt{s}$ separated into contributions of the $\Delta(1232)$, the higher mass $\Delta$ ($I=3/2$) and the $N^*$ ($I=1/2$) resonances in comparison to the parametrization \cite{teis}. The HADES results are superimposed as red symbols with error bars. The total pion production cross sections are equal to the sum of the resonance contributions listed in Table \ref{resonance_results}. For the isospin decomposition we have chosen cross sections of the selected resonances indicated in bold. Although the identification of resonances is ambiguous in the nucleon-pion invariant mass region of overlapping states, the decomposition is still feasible. It is performed by a comparison of the corresponding yields in the $n\pi^+$ and $p\pi^+$ invariant mass distributions for the $N^*$ and $\Delta$ resonances and is given as the product of the resonance cross section and the respective branching ratio. The comparison (see extracted values in the second column of Table \ref{resonance_results}) shows a qualitative agreement with the decomposition in \cite{teis} (third column). The differences are discussed below.

The $\Delta(1232)^+$ cross section obtained in our analysis is slightly higher than that of \cite{teis} and is closer to the cross section value used in GiBUU \cite{GiBUU}.  The total contribution of higher mass $\Delta$, with masses around  $M_{\Delta} \sim 1620$ MeV/c$^2$ and $M_{\Delta} \sim 1910$ MeV/c$^2$, is clearly larger in the fit \cite{teis} as compared to our results. One can hence conclude that the reduction of the respective cross sections applied in the  GiBUU version \cite{GiBUU} are in line with our findings. One can also notice that the cross sections for the higher mass $\Delta$ resonances are by a factor 2-3 larger in the UrQMD code \cite{urqmd} as compared to the GiBUU \cite{GiBUU} but lower for the $\Delta(1232)$.

For the $N^*$ resonances we can directly compare cross sections of $N(1520)$, $N(1535)$ and $N(1440)$. Our cross sections are closer to the values used in UrQMD \cite{urqmd}, except for $N(1535)$ which appears to be much larger in all models. As explained above, we fix the cross section for $N(1535)$ by the data on $\eta$ production. Although in  \cite{GiBUU} the sum of the cross sections for all $N^*$ resonances is similar to the model \cite{teis}, the relative partition is different, giving the largest weight to the $N(1440)$ and a smaller one to the $N(1675)$. One should also notice that the cross sections for $N(1720)$ and $N(1680$) used in \cite{urqmd} are also much higher by a factor of about $5-6$ than the ones used in \cite{GiBUU}. These cross sections, together with the cross section for the  $N(1520)$, $\Delta(1620)$ and $\Delta(1905)$ resonances play a major role for dielectron production because of their large $p\rho$ branching ratios.

\begin{figure}
\resizebox{0.5\textwidth}{0.24\textheight}{
  \includegraphics{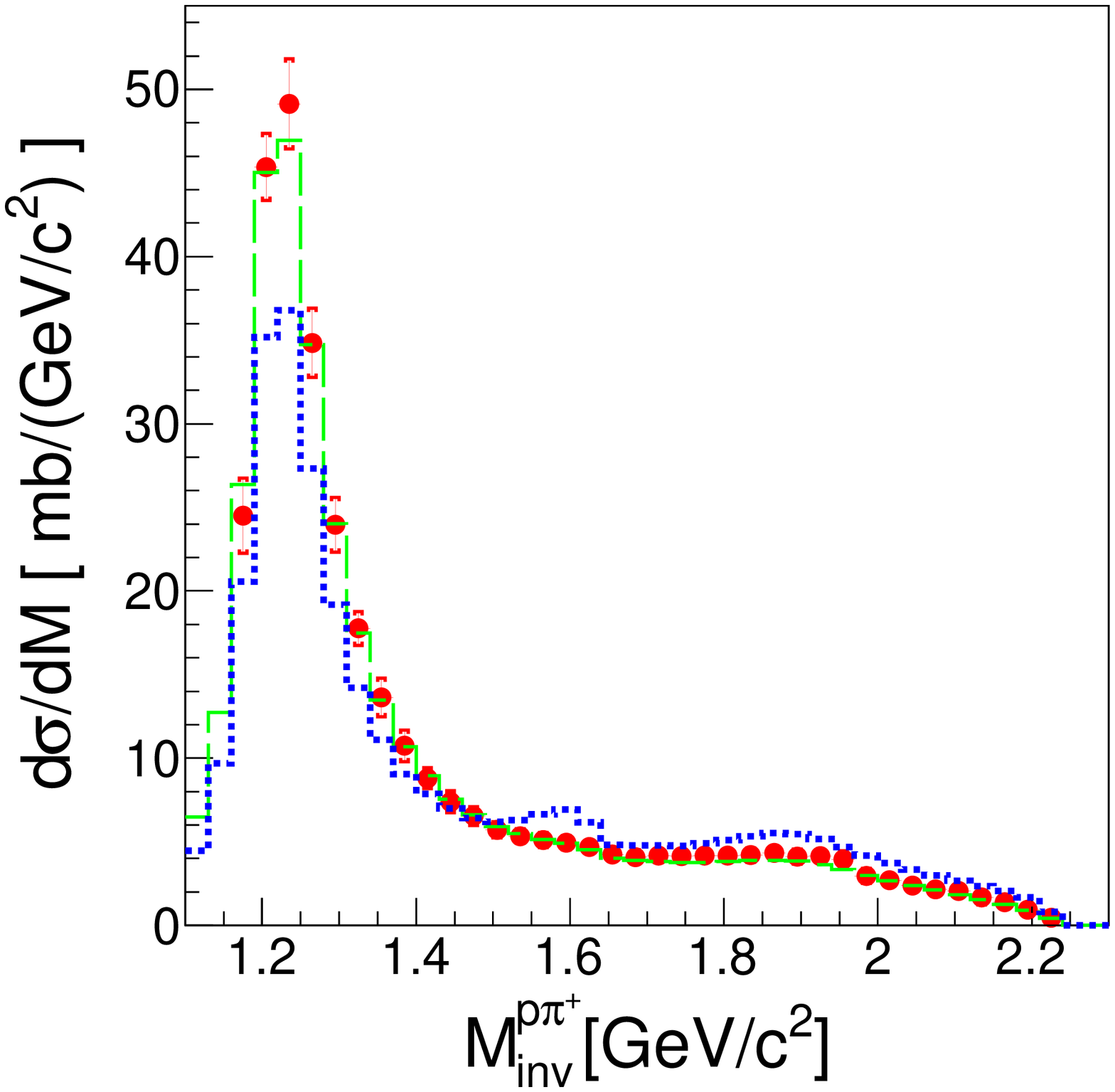}
  \includegraphics{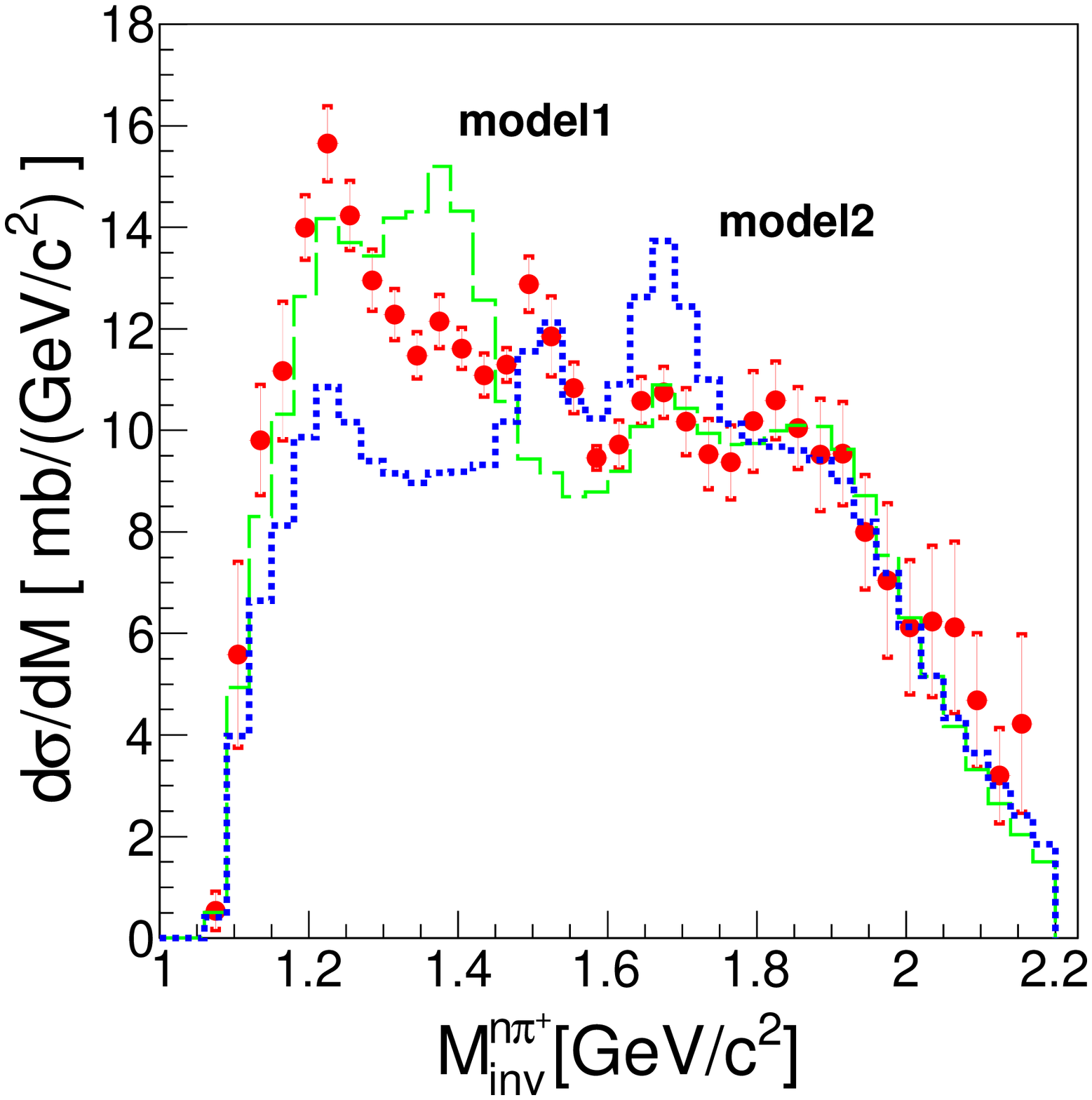}
}
\caption{$np\pi^+$ final state: Acceptance corrected $p\pi^+$ (left) and $n\pi^+$ (right) invariant mass distributions (symbols with error bars) compared to the simulation results using the resonance cross sections according to parametrizations taken from \cite{GiBUU} (dashed histogram - $model1$) or from \cite{urqmd} (dotted histogram - $model2$).}
\label{pnpi_mass_corr_transport}       
\end{figure}

The aforementioned features are visible in a comparison to the $pn\pi^+$ differential cross sections plotted as a function of the nucleon-pion invariant mass (Fig.~\ref{pnpi_mass_corr_transport}). The $p\pi^+$ invariant mass distribution is better described by simulations based on the cross sections used in \cite{GiBUU} (dashed histogram - $model1$). The parametrization used in \cite{urqmd} (dotted histogram - $model2$) underestimates the $\Delta(1232)$ production but overestimates the production of higher mass $\Delta$ states. On the other hand, the $n\pi^+$ invariant mass distribution, reflecting enhancements mainly due to the $N^*$ resonances, clearly shows that the strong $N(1440)$ production implemented in $model1$ is not supported by our data. There is also missing intensity around $N(1520)$ which could be explained by a larger resonance cross section, as deduced from our fit. Indeed, we have checked that taking the cross sections for both resonances and $N(1535)$ from our fit and leaving all the others without any change one can reproduce our result shown in Fig. \ref{pnpi_mass_corr}.

The comparison of the $n\pi^+$ invariant mass distribution to the calculations using the parametrization of resonance cross sections applied in $model2$ shows a clear overshoot in the mass region around $N(1680)$ / $N(1675)$ indicating too strong contributions from these resonances. On the other hand, the undershoot at low invariant masses is related to a too small $\Delta(1232)^{++}$ cross section.

\section{$\bf{ppe^+e^-}$ final state}
\label{ppee}

As described in Section \ref{channel_selection}, the $ppe^+e^-$ final state was selected by a cut on the $pe^+e^-$ missing mass $0.8$ GeV/$c^2$ $<M_{miss}^{pe^+e^-}<1.04$ GeV/$c^2$ (see Fig.~\ref{ppee_select}). This distribution and the $e^+e^-$ and the $pe^+e^-$ invariant mass distributions are used below in comparison to various models. All experimental distributions are normalized to the measured elastic scattering yields, and the simulation results are filtered through the acceptance and efficiency matrices followed by a smearing with the experimental resolution. The data are compared to simulations assuming the production cross sections $\sigma_{R}$ of baryon resonances from Table~\ref{resonance_results} and the $\omega$ and $\rho$ meson cross sections given in Section~\ref{sim}. These cross sections are converted to yields via the measured proton-proton elastic scattering yields of known cross section, as explained in Section~\ref{elastic}.

\subsection{Point-like $\bf{R N \gamma^*}$ coupling}
\label{section_qed}

We start with the assumption of a point-like $R N \gamma^*$ coupling, called hereafter "QED model", and the resulting baryon conversion yields given in \cite{zetenyi} which assumes constant eTFF.

\begin{figure*}
\resizebox{0.98\textwidth}{0.25\textheight}{
  \includegraphics{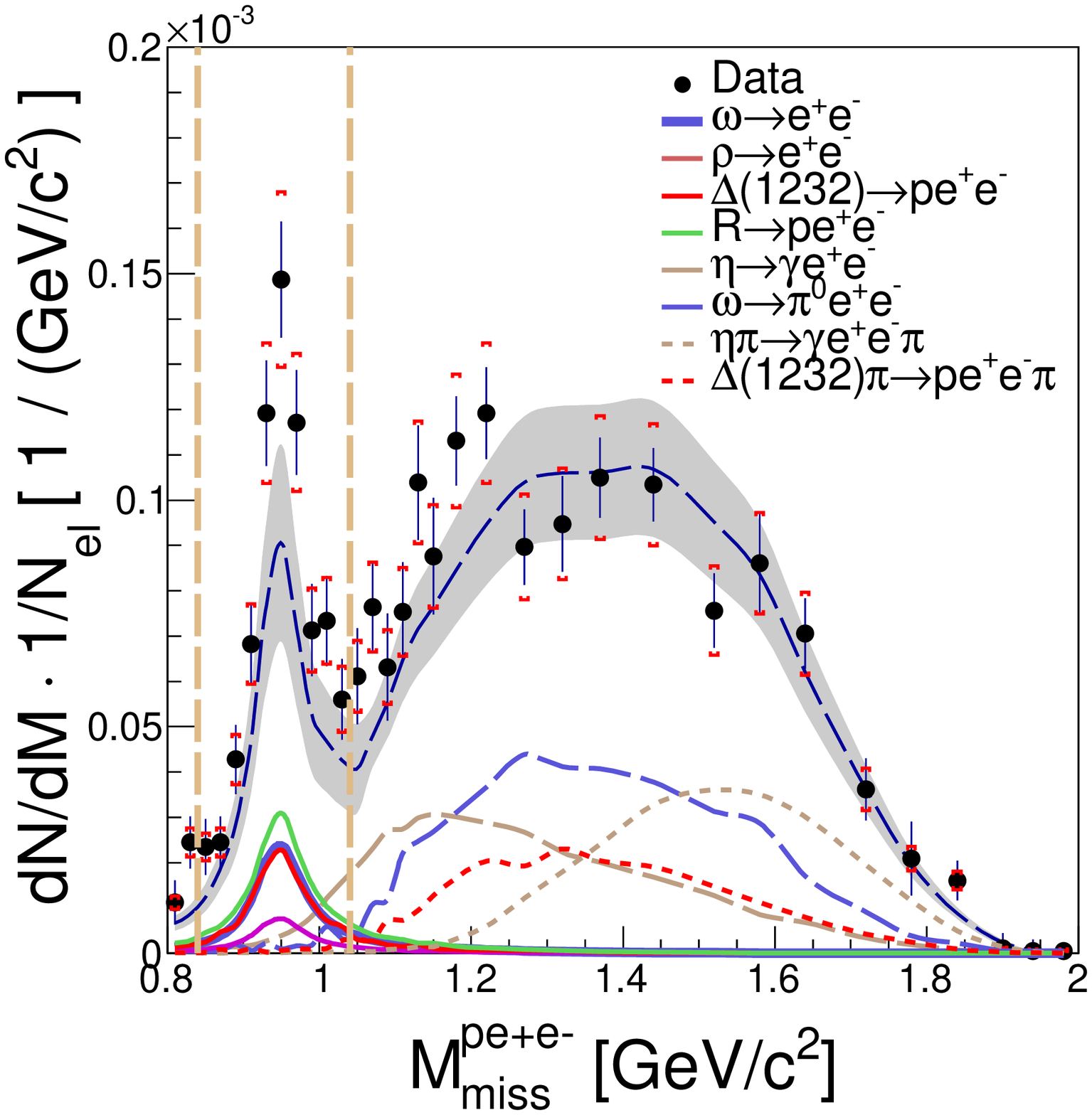}
  \includegraphics{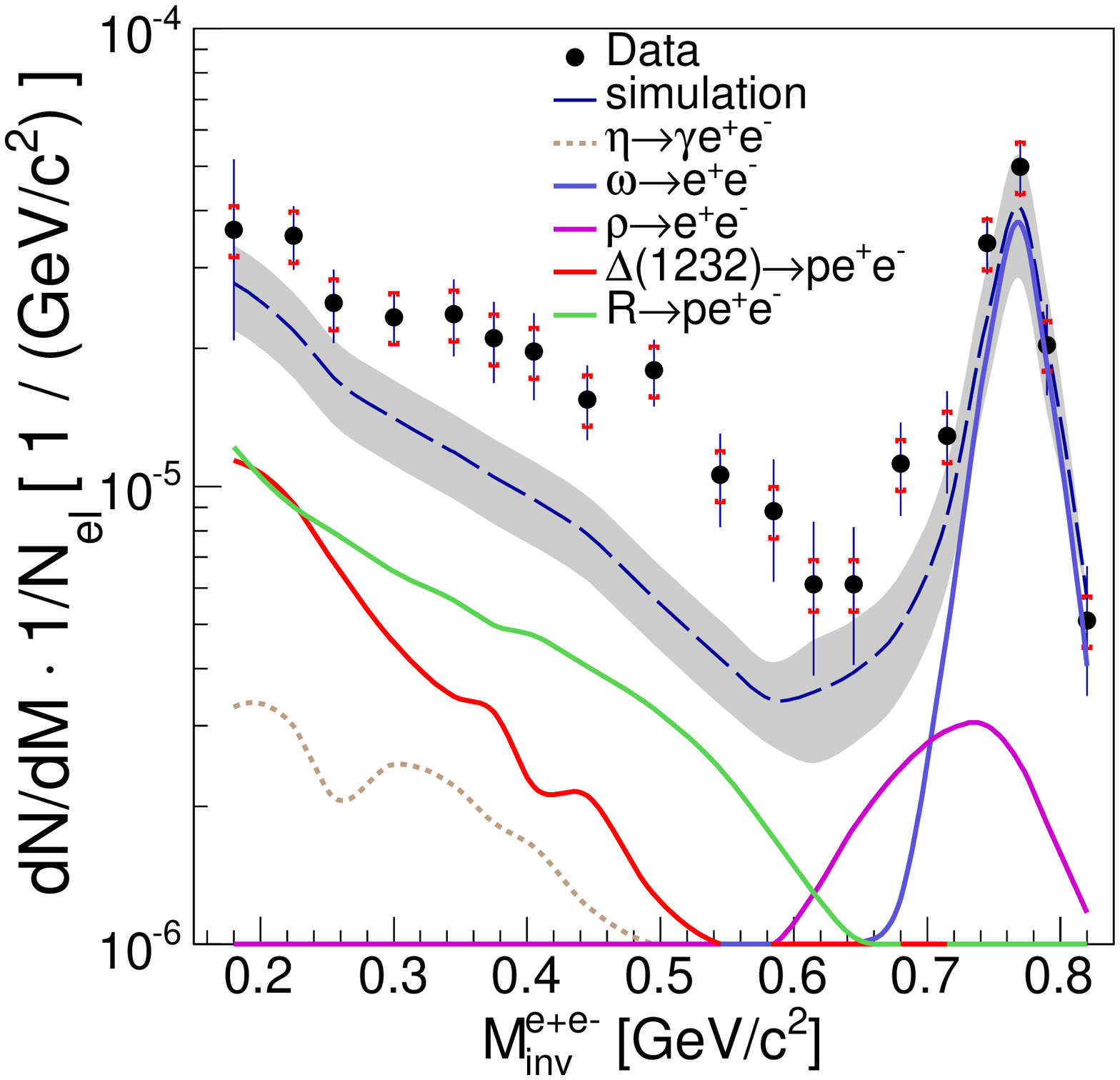}
  \includegraphics{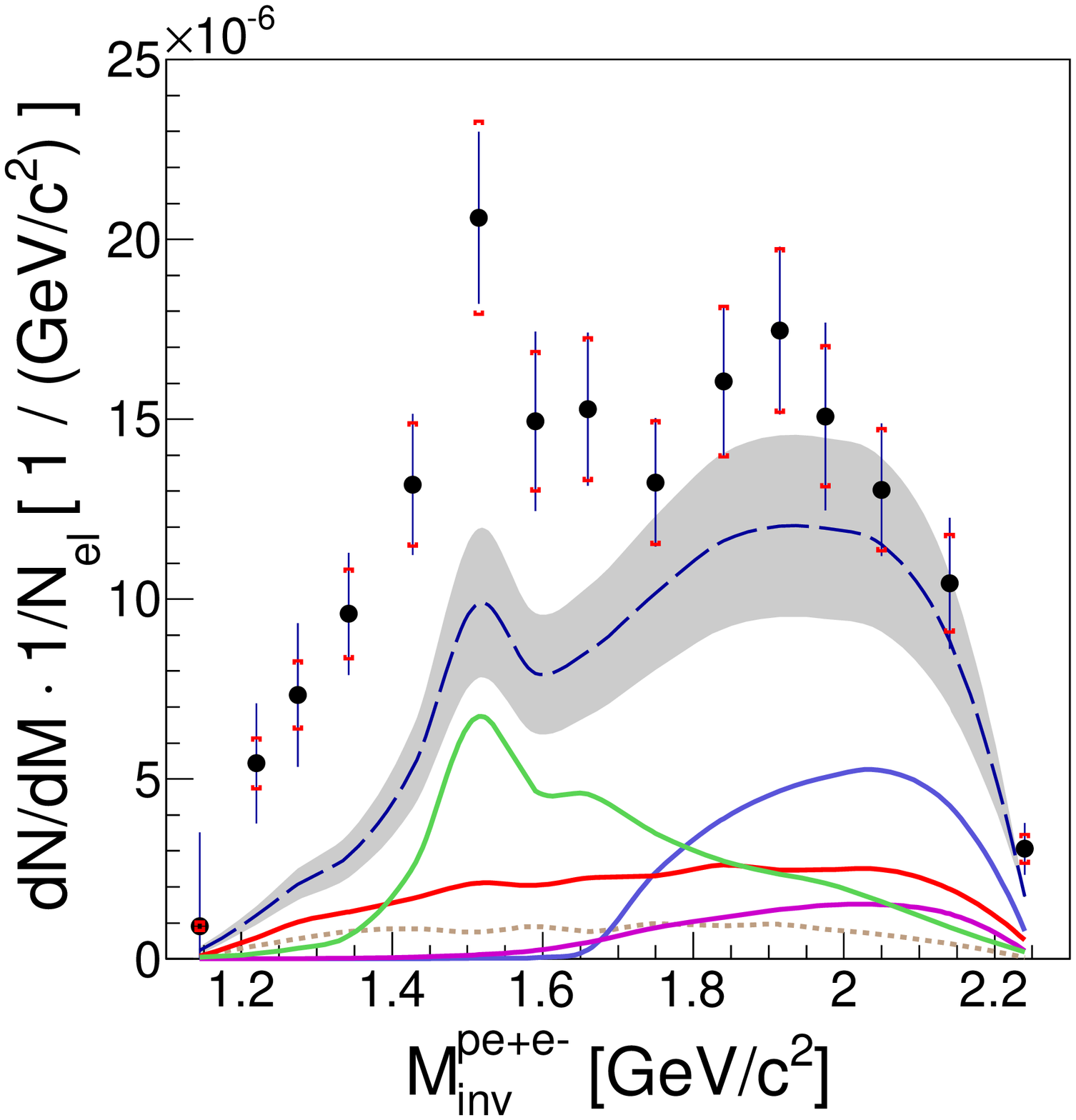}
}
\caption{$ppe^+e^-$ final state: $pe^+e^-$ missing mass (left), dielectron (middle) and $pe^+e^-$ (right) invariant mass distributions compared to the simulation result assuming a point-like $R N \gamma^*$ coupling ("QED-model"). The invariant mass distributions have been obtained for events inside the indicated window (vertical dashed lines in the left panel) on the $pe^+e^-$ missing mass. The hatched area indicates the model errors (for more details see text). Number of counts is per mass bin width.}
\label{ee_QED}       
\end{figure*}

The missing mass distribution of the $pe^+e^-$ system with respect to the beam-target system, after CB subtraction, is shown in Fig.~\ref{ee_QED} (left). The error bars represent statistical (vertical) and the normalization (horizontal) errors. The distribution is compared with the result of the simulation (dashed curve) including the baryon resonances and $\rho$, $\omega$ and $\eta$ meson sources. The baryon resonances included in the simulations are indicated by bold symbols in Table~\ref{resonance_results} and grouped into two contributions, appearing to be of similar size, originating from the $\Delta(1232)$ and the higher mass ($\Delta^+, N^*$) states. The hatched area uncovers the model uncertainties related to the errors of resonance and meson production cross sections (see below for a more detailed discussion).

In order to account for events with $M_{miss}^{pe^+e^-}>M_p$ the final states $p\Delta^{+,0}\pi^{0,+}$, $pp(n)\eta\pi^{0,+}$ were included in the simulations. Channels with two and more pions were omitted because of negligible contributions caused by smaller cross section and the small HADES acceptance for the very forward emitted protons. As one can see, a very good description of the $pe^+e^-$ missing mass distribution could be achieved with all the sources mentioned above, except for the yield in the proton missing-mass peak itself. It is important to note that the background under the proton peak, related to final states other than $ppe^+e^-$, is smaller than $6\%$. In particular, channels including the $\eta\rightarrow e^+e^-\gamma$ decay are strongly suppressed.

The middle part of Fig.~\ref{ee_QED} displays the $e^+e^-$ invariant mass distribution for the events within the $pe^+e^-$ missing mass window, shown by the vertical dashed lines in the left pannel. It is compared to the simulation including dielectron sources originating from the baryon resonance decays and the two-body meson $\rho$, $\omega\rightarrow e^+e^-$ decays. As one can see, a very good agreement in the vector mass pole is achieved. Since the exclusive production cross section of vector mesons at this energy are rather well known, the agreement confirms that the normalization and the simulations of the HADES acceptance and reconstruction efficiencies are under control. On the other hand, an excess of the contributions from the baryon resonances is clearly visible below the vector meson pole. The effect is obviously related to the apparent excess in the proton missing-mass window. This is, however, not a surprise because one expects contributions from off-shell couplings of the resonances to the vector mesons. As discussed above, it is expected that such couplings modify the respective eTFF which were assumed to be constant in the simulations. Therefore, the observed enhancement below the vector mass pole can be interpreted as a fingerprint of the anticipated contribution.

The hatched area presents the model error on the dielectron conversion yields related to the discussed ambiguities of the resonance assignments. Apart from the resonance production cross sections, the overlapping states differ also in the branching ratios for the Dalitz decay (see Tables \ref{resonance_table} and \ref{resonance_results}). However, the effect on the pair yield  (hatched area) turns out to be rather moderate. This is because the relative variation of the pair yield due to changes in the resonance production cross sections is compensated by the respective changes in the branching ratios for the dielectron conversion.  Consequently, one can conclude that the excess above the calculated yield cannot be explained by another choice of the resonances in our calculations. The substantially different shape of the experimental invariant mass distribution, as compared to the simulation, indicates also the importance of the off-shell vector couplings.

This conclusion seems to be corroborated by the comparison of the $pe^+e^-$ invariant mass distribution with the simulation, displayed in Fig.~\ref{ee_QED} (right), which shows that the excess is indeed located around the $N(1520)$ resonance known to have a sizable decay branch to the $\rho$ meson.

\subsection{Comparisons to models assuming a "full" resonance-$\bf{\rho}$ coupling scheme}

In this subsection we present a comparison of the $e^+e^-$ and $pe^+e^-$ invariant mass distributions from our experiment to the results of calculations assuming dielectron production through the resonance decay $R \rightarrow p\rho \rightarrow p e^+e^-$. As already mentioned, such a factorization scheme is used in transport  models like the GiBUU and the UrQMD. The results of the two models were recently published \cite{GiBUU,urqmd1} and were compared to our inclusive data \cite{pp35GeV}. In order to compare the calculations of the contributions to the exclusive $ppe^+e^-$ channel we have to select only final states including single resonance production. The respective cross sections are given in Table~\ref{resonance_results} and the branching ratios to $p\rho$ are listed in \cite{GiBUU} and \cite{urqmd1}. Table~\ref{resro} summarizes these branching ratios (columns "GiBUU" and "UrQMD") together with more recent results from a multichannel partial wave analysis which are discussed below in this section.

\begin{table}
\begin{tabular}{|ccc|ccc|}
  \hline
  Resonances & GiBUU & UrQMD & KSU & BG & CLAS\\
  \hline
  $N(1520)$   &  $21$ & $15$ & $20.9(7)$   & $10(3)$  & 13(4)\\
  $\Delta(1620)$  &  $29$  & $5$  & $26(2)$    & $12(9)$  & 16 \\
  $N(1720)$   &  $87$ & $73$ & $1.4(5)$    & $10(13)$ &  - \\
  $\Delta(1905)$  &  $87$ & $80$ & $<14$       & $42(8)$  &  - \\
  \hline
\end{tabular}
\caption{\small Branching ratios (in percent) for $R \rightarrow N\rho$ decays applied in GiBUU \cite{GiBUU} (second column) and UrQMD \cite{urqmd1} (third column) for the most important dielectron sources. KSU: $BR(N\rho)$ and its error (in brackets) from multichannel PWA \cite{shrestha}, BG: the difference between the total and the sum of all determined partial branching ratios (except $N\rho$) from the Bonn-Gatchina group \cite{anisovich}. CLAS: results from the analysis \cite{mokeev}. For more details see the text.}
\label{resro}
\end{table}

We start with the GiBUU events, provided by the authors of \cite{GiBUU}, which were filtered through the HADES acceptance and reconstruction efficiency matrices. For the resonance production a non isotropic production was assumed according to the measured $t$ distributions presented in Section \ref{hadrons_hades}. The $\omega$ meson production is generated assuming uniform phase space population.

\begin{figure}
\resizebox{0.5\textwidth}{0.25\textheight}{
  \includegraphics{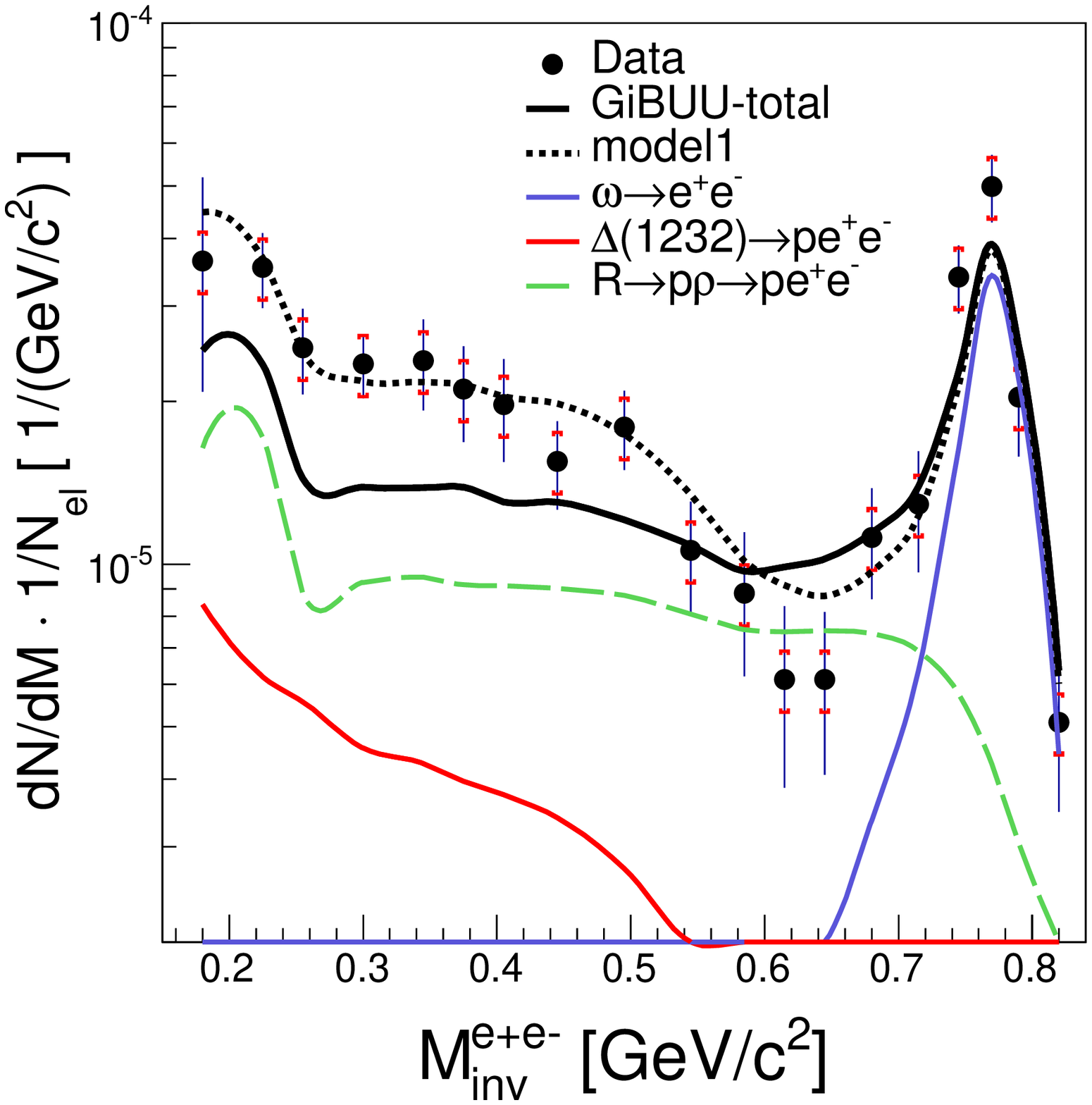}
  \includegraphics{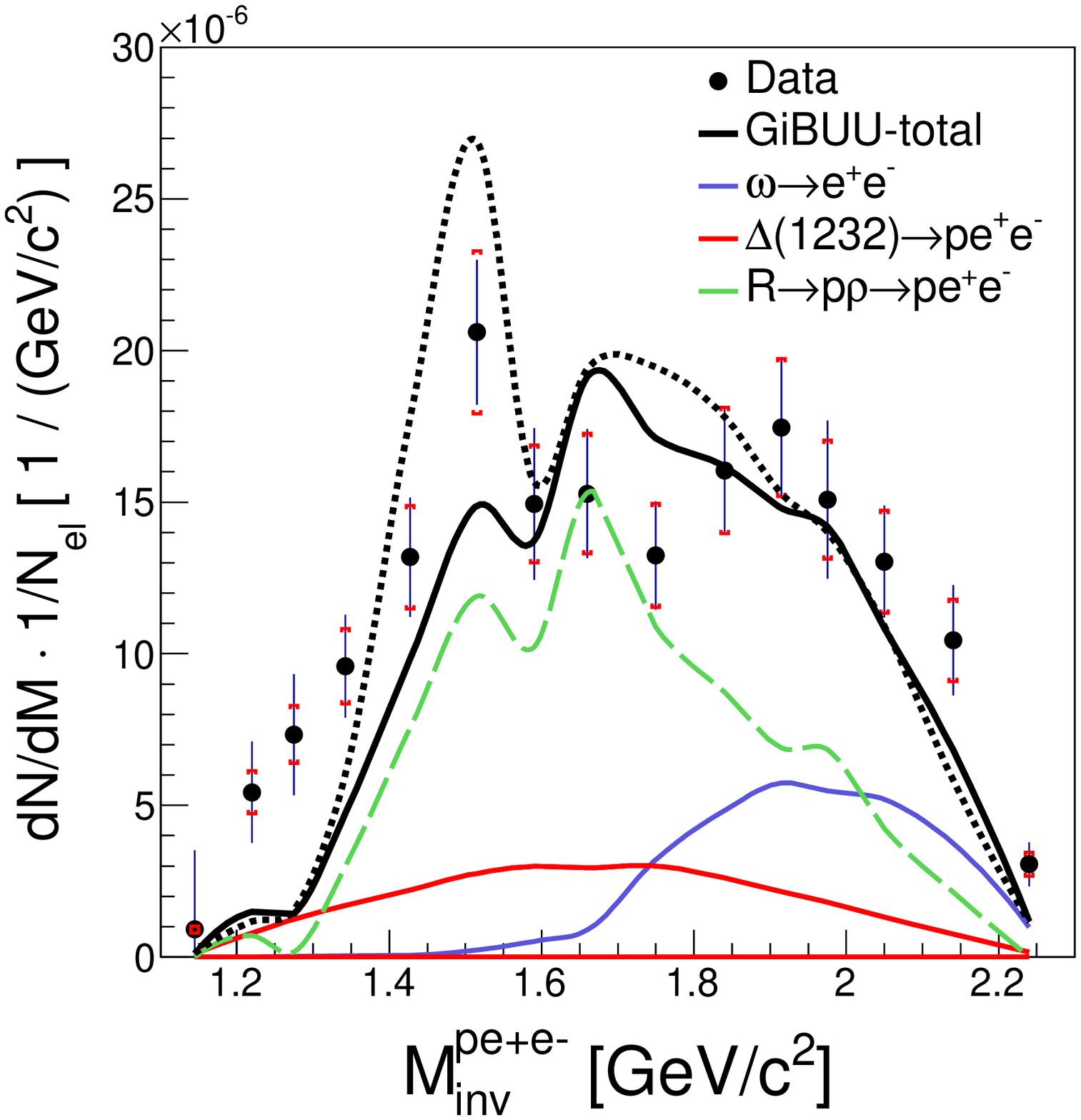}
}
\caption{Experimental dielectron (left) and $pe^+e^-$ (right) invariant-mass distributions compared to simulations based on the input from GiBUU (solid curve). Contributions from higher mass resonances, $\Delta(1232)$ and the $\omega$ meson are indicated separately. Dotted curves show results of calculations using modified cross sections and $R\rightarrow N\rho$ branching ratios from \cite{anisovich}. Number of counts is per mass bin width. For details see the text.}
\label{transport}       
\end{figure}

The two plots in Fig.~\ref{transport} show a comparison of the dielectron and the $pe^+e^-$ invariant mass distributions to the results of calculations normalized to the same elastic scattering yield. The total yield (solid curves) is decomposed into the contributions originating from the $\Delta(1232)$ (red curves), the $\omega$ meson (blue curves) and the higher mass resonances (dashed green curve) which are mainly the decays of $N(1520)$ $(38\%)$, $N(1720)$ $(22\%)$, $\Delta(1620)$ $(15\%)$ and $\Delta(1905)$ $(6.5\%)$. The measured distributions are well described, except some lacking intensity at low dielectron and $pe^+e^-$ invariant masses and some overshoot just below the vector meson pole. The missing yield might suggest an even stronger contribution of $N(1520)$, as also indicated by the comparison to pion spectra in Fig.~\ref{pnpi_mass_corr_transport}, where the calculations based on cross sections used in the GiBUU ($model1$) do not describe the $n\pi^+$ invariant mass distributions around $1.5$ GeV/c$^2$. On the other hand, an application of the cross section for $N(1520)$ and $N(1440)$ obtained from our analysis would overestimate the measured dielectron yield almost by a factor $2$.

Since the resonance sources contributing to the dielectron production in UrQMD \cite{urqmd} are almost the same as in GiBUU \cite{GiBUU}, one can estimate the corresponding yields. Indeed, according to \cite{urqmd1} (see figure~7 in there) the main contributions to the $\rho$ production stem from $N(1720)$, $N(1520)$, $\Delta(1905)$ and $N(1680)$, respectively. The production cross sections are given in Table \ref{resonance_results} and are by a factor $5-6$ larger than the corresponding cross sections used in the GiBUU code \cite{GiBUU}. Consequently, the calculated total dielectron yield below the vector meson pole, including the $\Delta(1232)$ contribution, is overestimated by a factor of about 3. Also the authors of \cite{urqmd1} came to similar conclusions comparing their calculations to the inclusive dielectron production measured by DLS \cite{dls}. The UrQMD code is recently under revision and we hope that our data on exclusive channels will help to improve the description of dielectron production.

From the presented comparison one can see that, although both models were well tuned to describe the total pion production cross sections, the predictions for dielectron production differ substantially. This is not a surprise since, in spite of the large branching ratios for the $N\rho$ decays assumed in the calculations, dielectrons are very sensitive to the resonance contributions. In particular, $e^+e^-$ contributions from Dalitz decays of higher mass resonances are significant, larger than expected from $\Delta(1232)$ Dalitz decay, and require a good understanding of the $R\rightarrow pe^+e^-$ decay mechanism. In the factorization scheme, with off-shell $\rho$-resonance coupling, the dielectron yield depends on the $R\rightarrow p\rho$ branching ratios which are taken in both models within the limits given by the PDG \cite{pdg}. The extracted parameters are based on various multichannel analyses of pion induced reactions (mainly two-pion production), suffering from low statistics. A new comprehensive multichannel analysis of the pion and photon induced reactions, performed by Shrestha and Manley (KSU) \cite{shrestha} and by the Bonn-Gatchina (BG) group \cite{anisovich}, however, shows smaller branching ratios for the $N\rho$ decays (see Table \ref{resro}). In the BG analysis the dominant channel for the two-pion production is the $\Delta \pi$ channel. The group does not provide any branching ratios for the $p\rho$ decay ($\pi^+\pi^-$ final state is not included in the analysis), however, from the provided branching ratios (mainly $\pi N$ and $\Delta \pi$) one can estimate the contribution left for the $p\rho$ decay. Table \ref{resro} shows the respective estimates, which for the most important resonances $N(1520)$, $N(1720)$ $\Delta(1620)$ predict branching ratios of the order of $10\%$ only. Also the recent results from CLAS \cite{mokeev} suggest lower values of the branching ratios (see the rightmost column in Table \ref{resonance_results}).

Using the BG branching ratio would lead to an underestimation of the dielectron yield if the cross sections applied in GiBUU \cite{GiBUU} are strictly used. However, if the higher cross sections for the $N(1520)$ and smaller for the $N(1440)$, $N(1535)$, as extracted from our simulations, are taken the calculation explains the measured $ppe^+e^-$ yield slightly better, as seen in Fig. \ref{transport} (model1 - dashed dotted curve). Hence, it remains still a subject of future work, both on theoretical and experimental sides, to better constrain the properties of the $R\rightarrow pe^+e^-$ decay. In this context, future experiments of HADES with pion beams aiming at investigations of pion and dielectron production in the second resonance region are expected to provide new valuable information.

 \subsection{Summary and Outlook}
\label{outlook}
We have presented a combined analysis of the three exclusive channels $pp\pi^0$, $pn\pi^+$ and $ppe^+e^-$ in $p+p$ collisions using a proton beam with a kinetic energy of 3.5~GeV ($\sqrt{s}=3.18$ GeV). From the pion production channels we have estimated exclusive $\Delta$ and $N^*$ resonance production cross sections by  means of a resonance model. We have also derived empirical angular distributions for the production of resonances showing a strong forward-backward peaking which is characteristic for peripheral reactions. A good description of the experimental data in the detector acceptance has been achieved allowing for an extrapolation to the full solid angle and an extraction of the pion production cross sections. Although the applied model assumes a simplified reaction mechanism ignoring interferences between various intermediate states it describes the data surprisingly well. Further studies, e.g. by means of the partial wave analysis, are on the way, including also data on lower energy, to estimate the effect of the latter and to study production of resonances in more detail. Nevertheless, the obtained results are very useful for a comparison of various parameterizations of the production of resonances used in the transport codes, as shown for the GiBUU and UrQMD codes.

Dielectron production from electromagnetic baryon-resonance Dalitz-decays and two-body $\omega$ meson decay ($\omega\rightarrow e^+e^-$) have been investigated in the $ppe^+e^-$ channel. Clear signals of the $\omega$ meson and the resonance decays have been established. In particular, a significant yield below the vector meson pole has been measured and attributed to the Dalitz decays of baryon resonances. Using the resonance model approach, upper limits for the various resonance contributions to the dielectron spectrum have been obtained assuming point-like baryon-virtual-photon couplings. The calculated dielectron yields cannot reproduce the measured yield and suggest strong off-shell vector meson couplings, which should influence the respective electromagnetic Transition Form-Factors (eTFF). Upcoming  theoretical studies of the eTFF in the time-like region are eagerly awaited for a more detailed comparison with our data.

An alternative approach for the Dalitz decay of resonances assuming a factorization scheme $R\rightarrow p\rho\rightarrow p e^+e^-$ was studied following the implementation used in the GiBUU and UrQMD codes. The GiBUU calculations explain the dielectron and $pe^+e^-$ invariant mass distributions, except the low-mass region which are due to a too small $N(1520)$ contribution visible also in the comparison of the model to the $n\pi^+$ invariant mass distribution. On the other hand simulations based on the resonance cross sections used in UrQMD overestimate dielectron yields by a factor 3. However, the calculated dielectron yields depend strongly on the $R\rightarrow p\rho$ branching ratios which, according to new results from multichannel analyses of pion and photon reactions off the proton, might be smaller than presently used in transport calculations. This conclusion is also corroborated by our model calculations employing smaller branching ratios and the cross sections for resonance production derived from the $pp\pi^0$ and $np\pi^+$ channels. Further theoretical studies, including our results on exclusive $ppe^+e^-$, are needed to better understand the electromagnetic decays of baryon resonances. In this respect, pion-proton collisions with simultaneous reconstruction of different final meson states are promising to pin down the excitation of resonances and couplings to virtual photos.

\section{Acknowledgements}

We would like to thank A.V. Sarantsev, J. Weil, G. Wolf and M. Zetenyi for stimulating discussions and
valuable remarks. In particular we would like to thank J. Weil for providing us events from the GiBUU code.

The collaboration  is very thankful to the GSI/SIS18 accelerator stuff for providing us en excellent beam. The collaboration gratefully acknowledges support by LIP Coimbra, Coimbra (Portugal) PTDC / FIS / 113339 / 2009, SIP JUC Cracow, Cracow (Poland) 2013/10/M/ST2/00042, HZ Dresden-Rossendorf (HZD), Dresden (Germany) BMBF 06DR9059D, TU M\"{u}nchen, Garching (Germany) MLL M\"{u}nchen DFG EClust 153 VH - NG - 330 BMBF 06MT9156 TP5 GSI TMKrue 1012 NPI AS CR, Rez, Rez (Czech Republic) MSMT LC07050 GAASCR IAA100480803 USC - S. de Compostela, Santiago de Compostela (Spain) CPAN: CSD2007 - 00042 Goethe-University, Frankfurt (Germany) HA216 / EMMI HIC for FAIR (LOEWE) BMBF: 06FY9100I GSI.

\section{Appendix}

\begin{figure*}
\hspace{-0.5cm}
\resizebox{1.\textwidth}{0.22\textheight}{
  \includegraphics{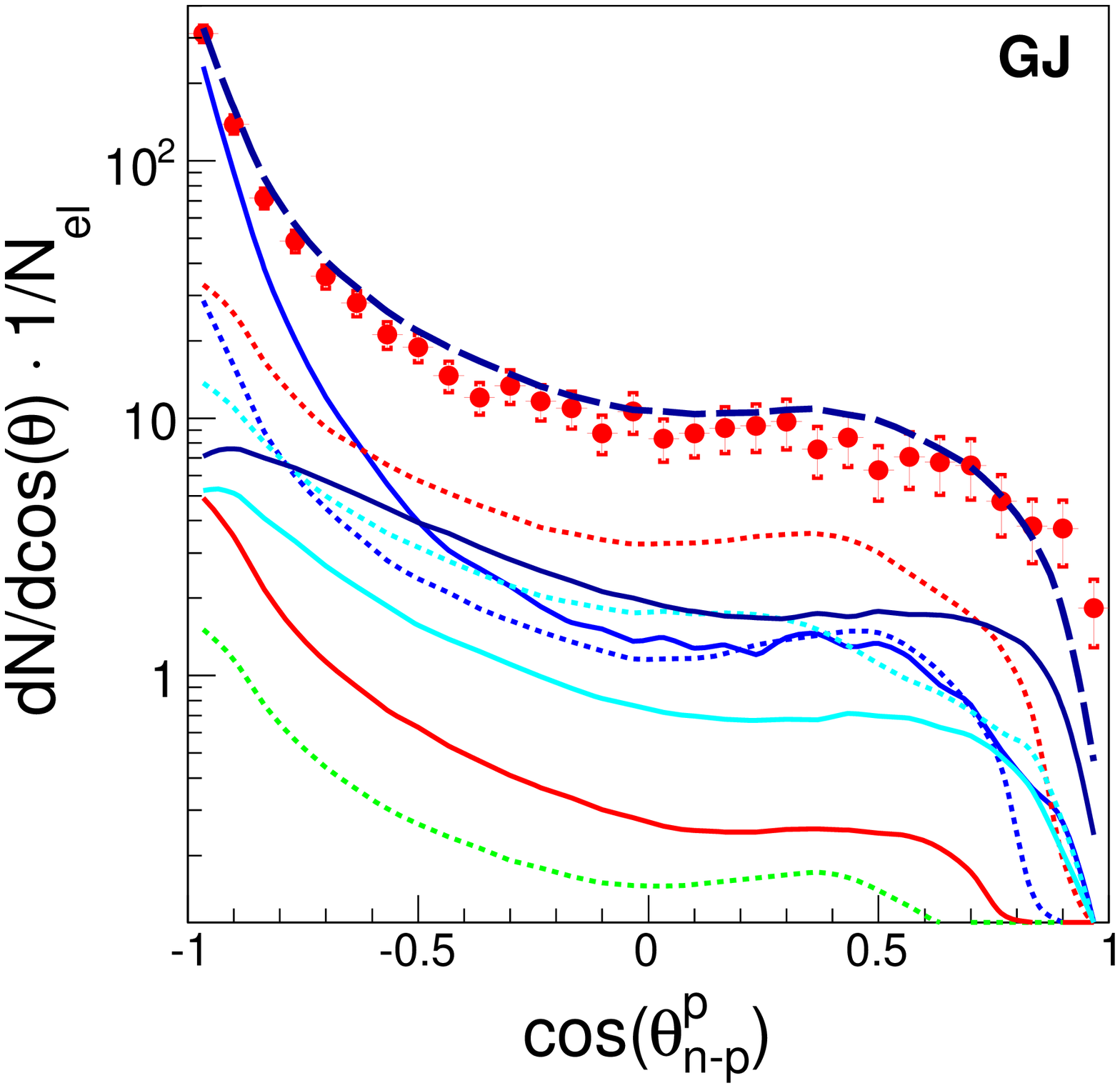}
  \includegraphics{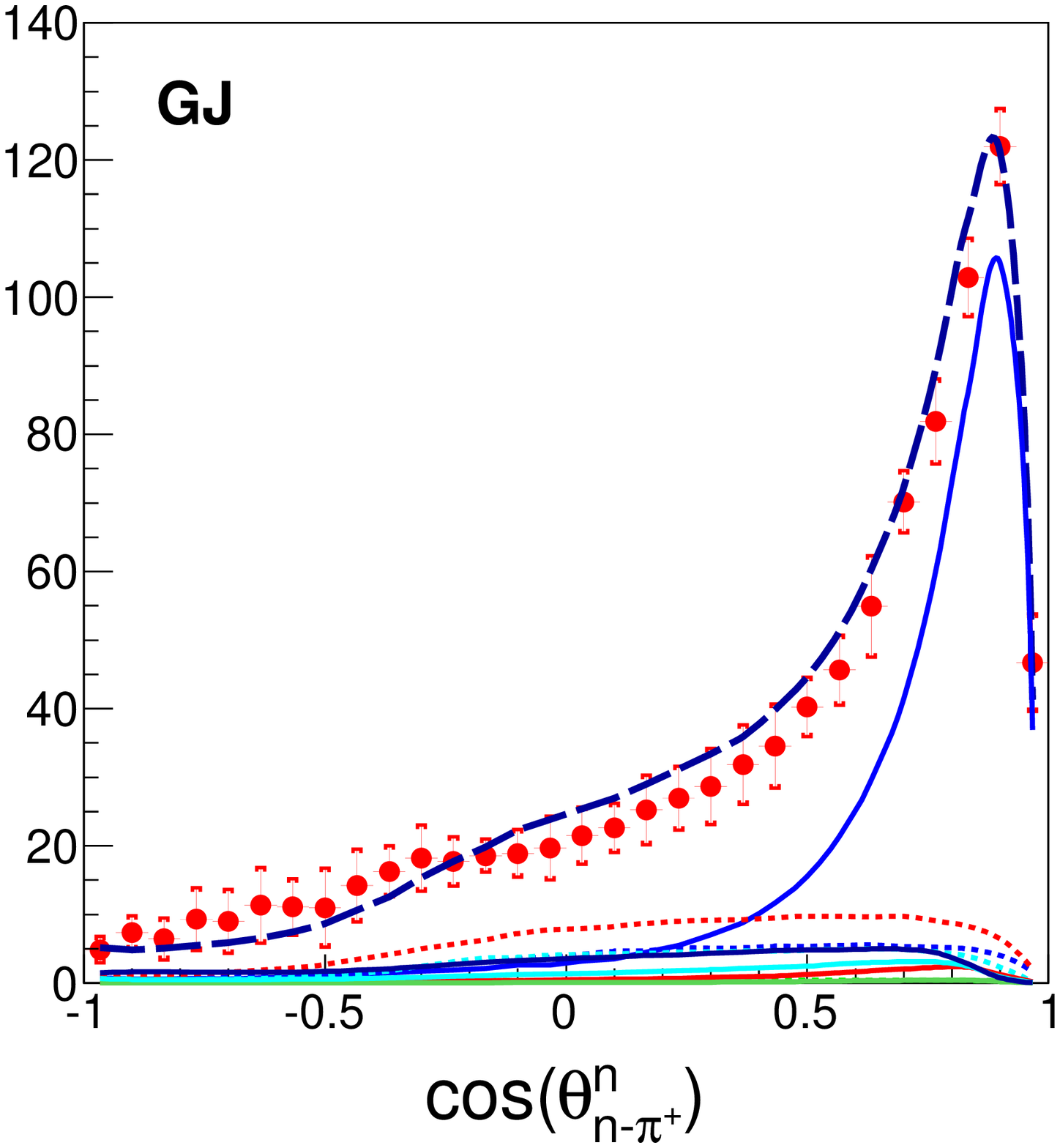}
  \includegraphics{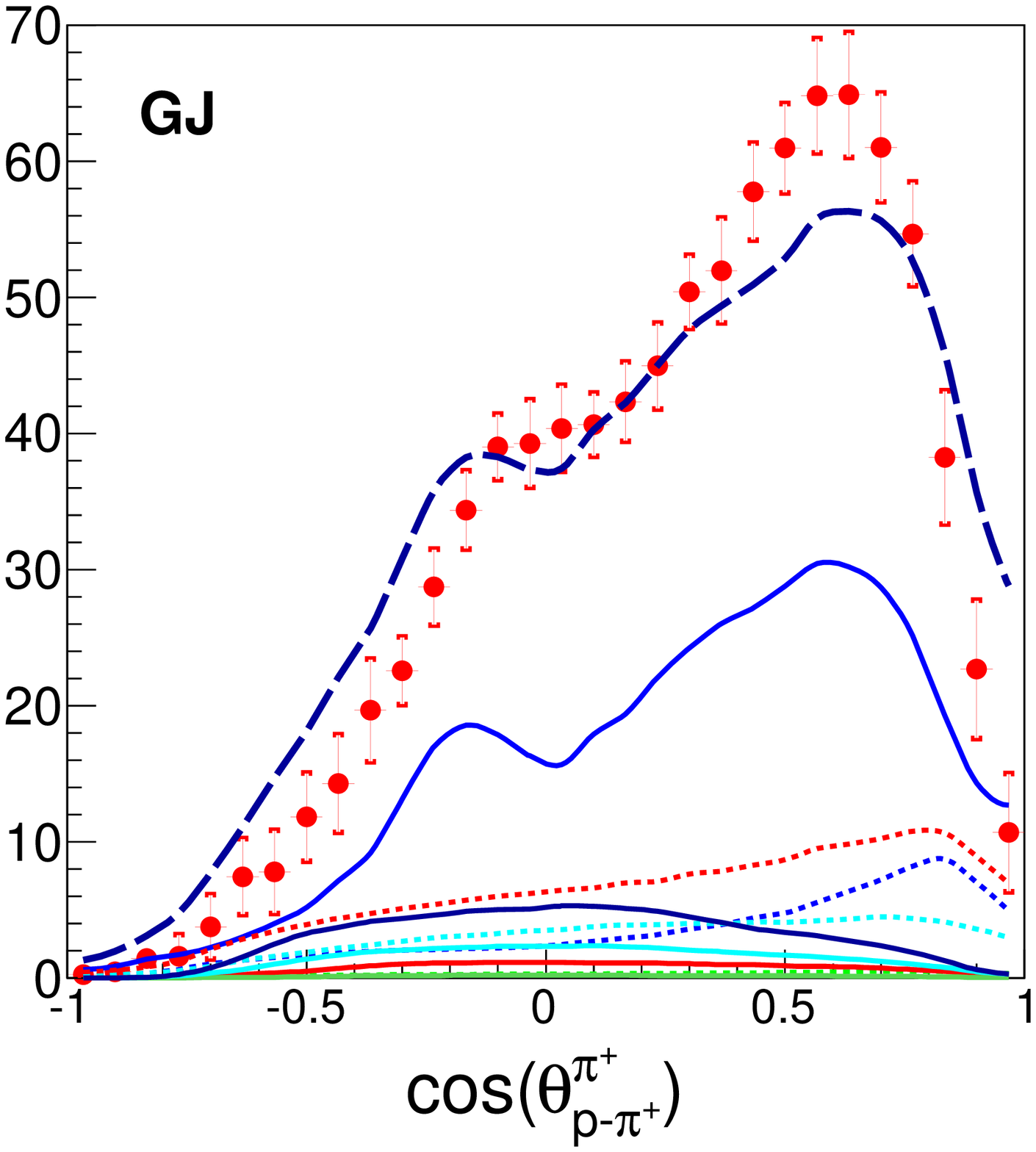}
  }
  \resizebox{1.\textwidth}{0.23\textheight}{
  \includegraphics{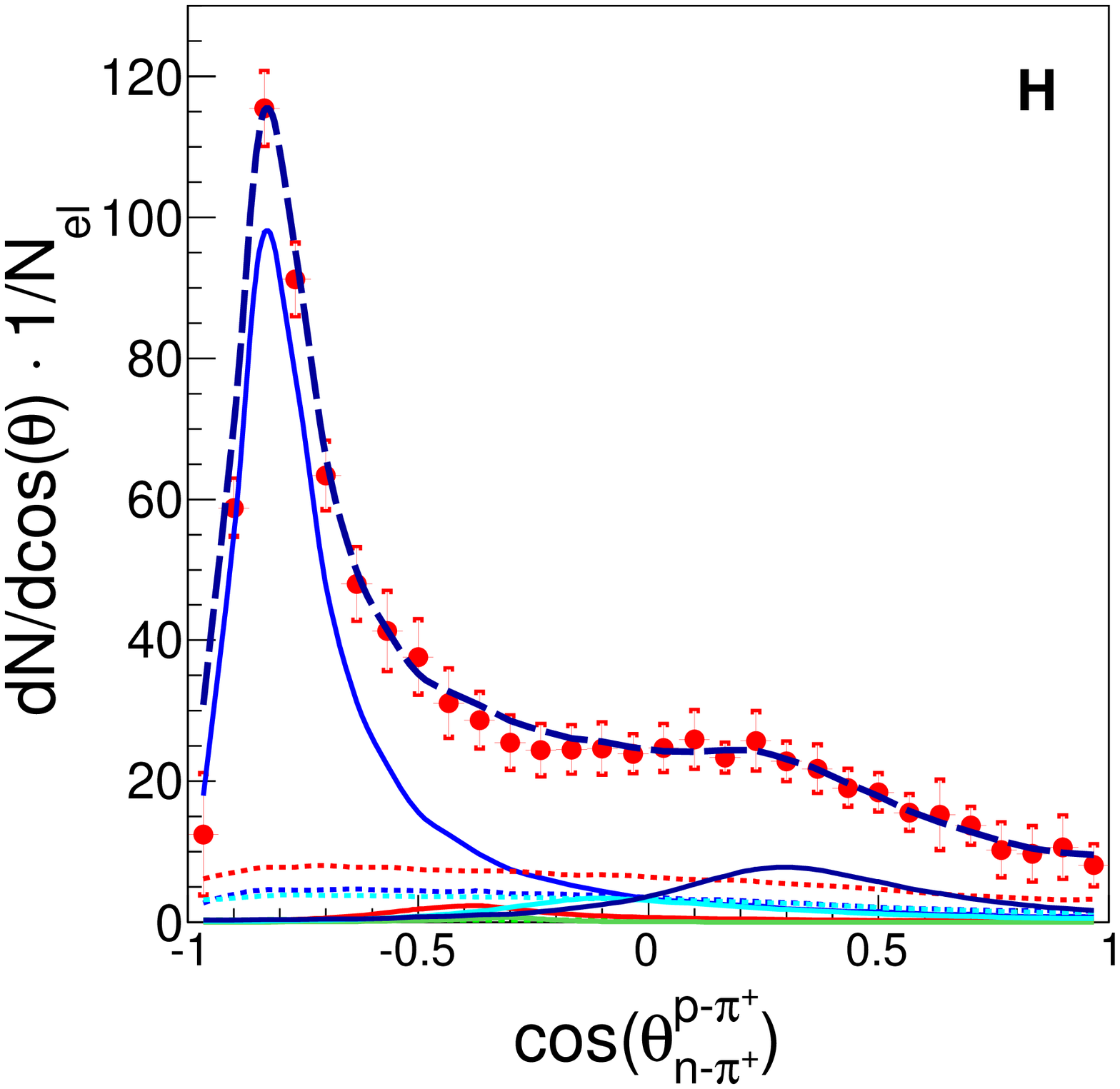}
  \includegraphics{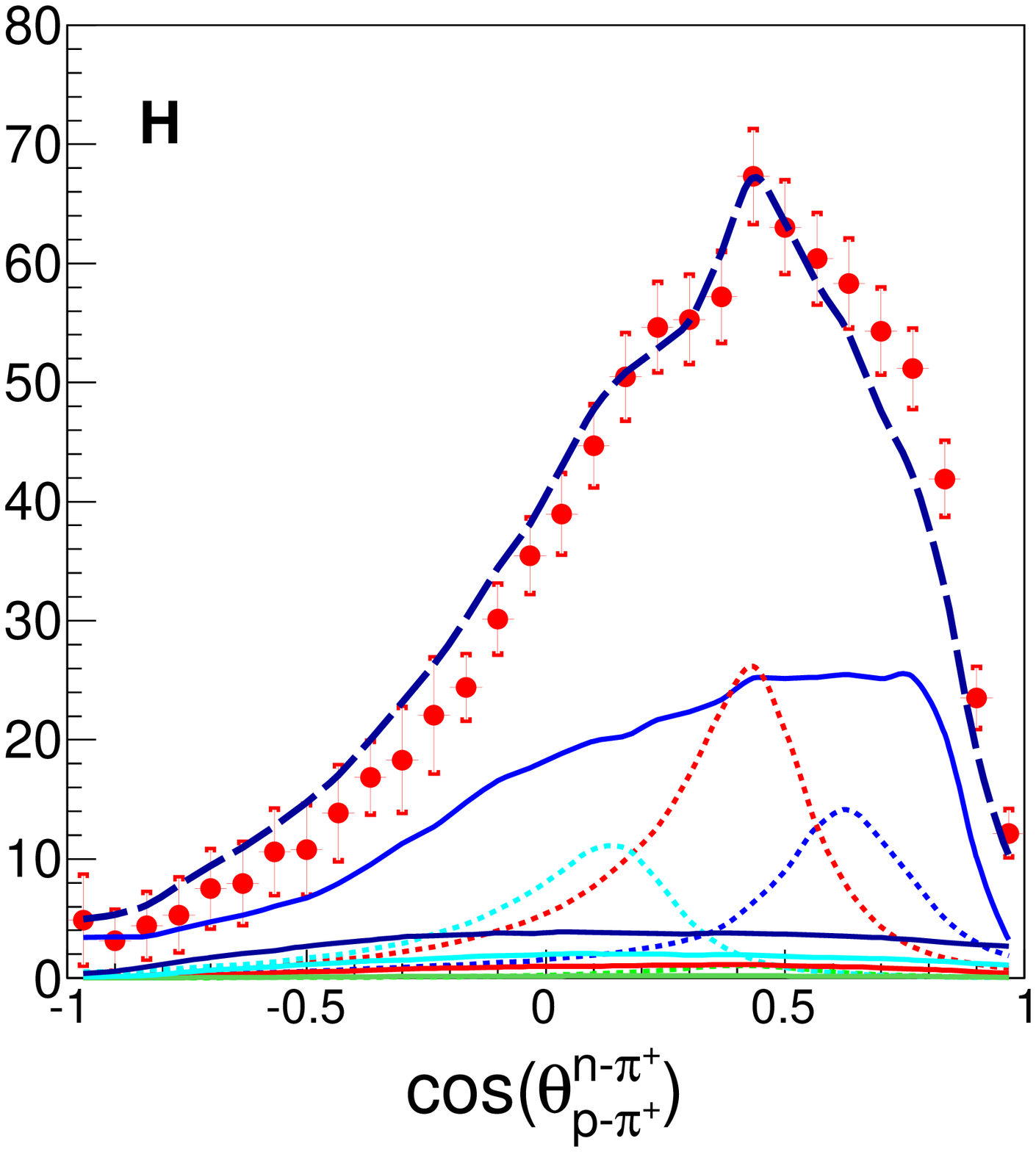}
  \includegraphics{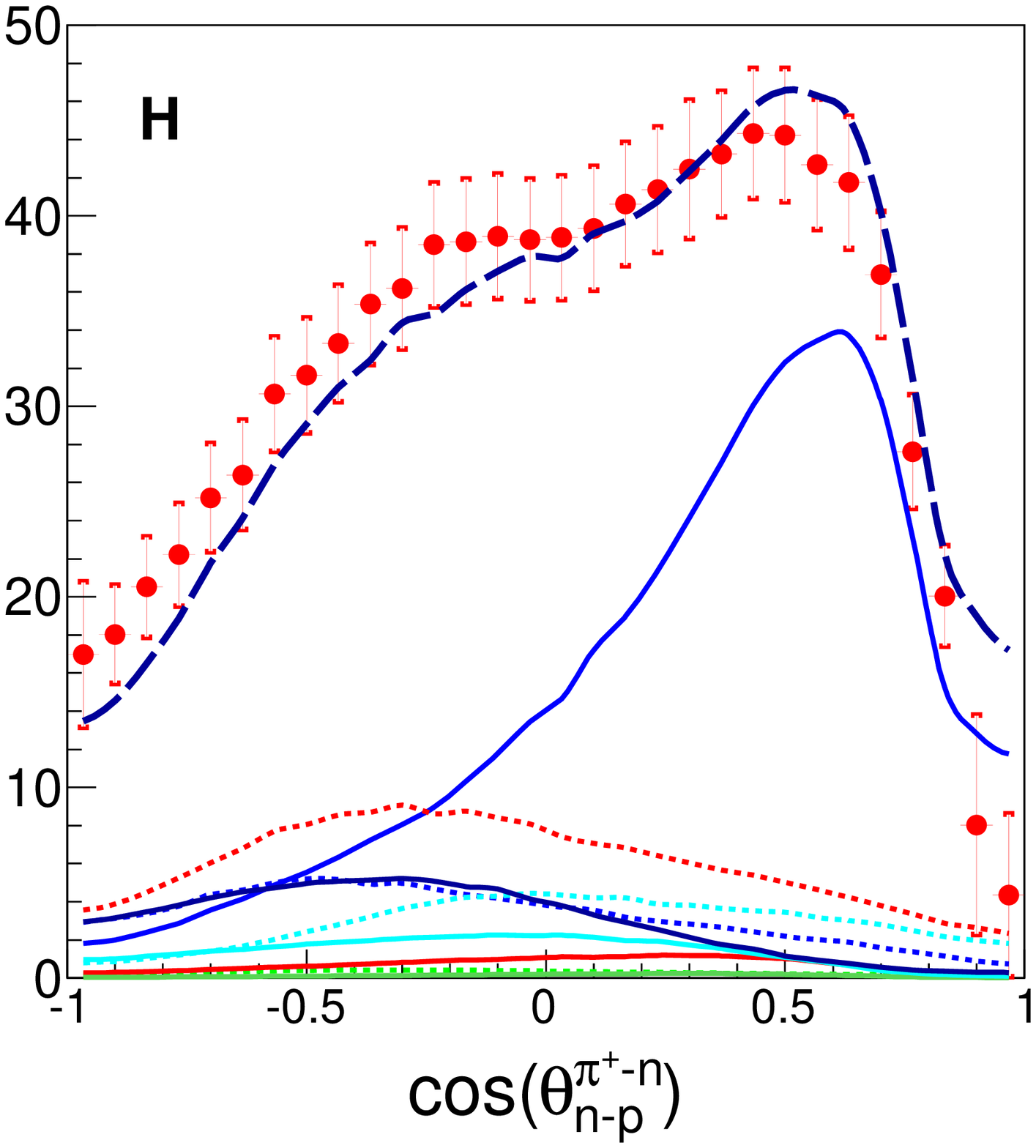}
  }
 \caption{Angular distributions in the Gottfried-Jackson reference frame (top) for the $pn\pi^+$ final state compared to the results of simulations (dashed curve) decomposed into contributions of various resonances and in the helicity (bottom) reference frame (for the line style, see Fig. \ref{pnpi_mass}).}
\label{GJH_pnpi1}       
\end{figure*}

\begin{figure*}
\resizebox{1.0\textwidth}{0.22\textheight}{
  \includegraphics{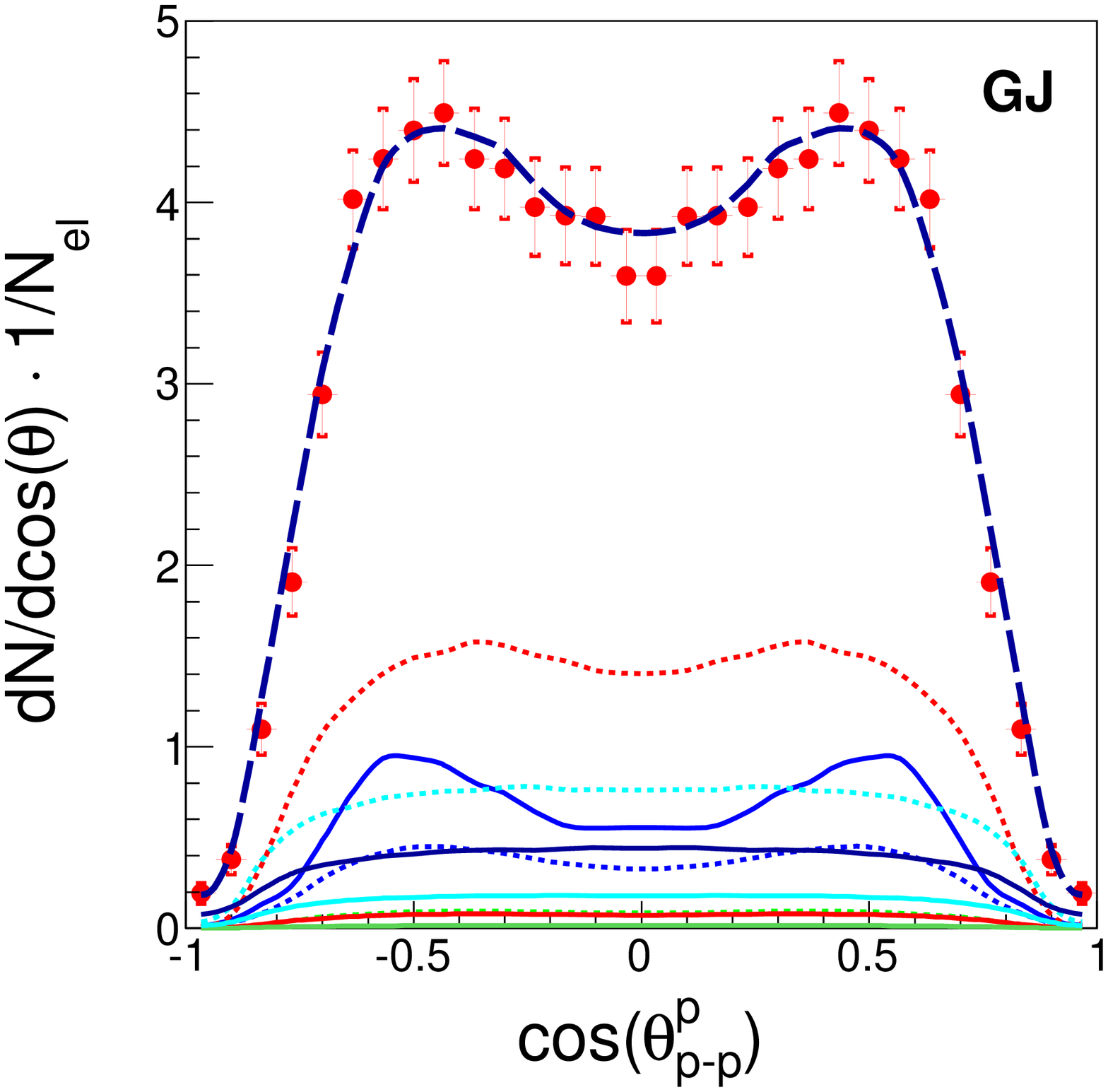}
  \includegraphics{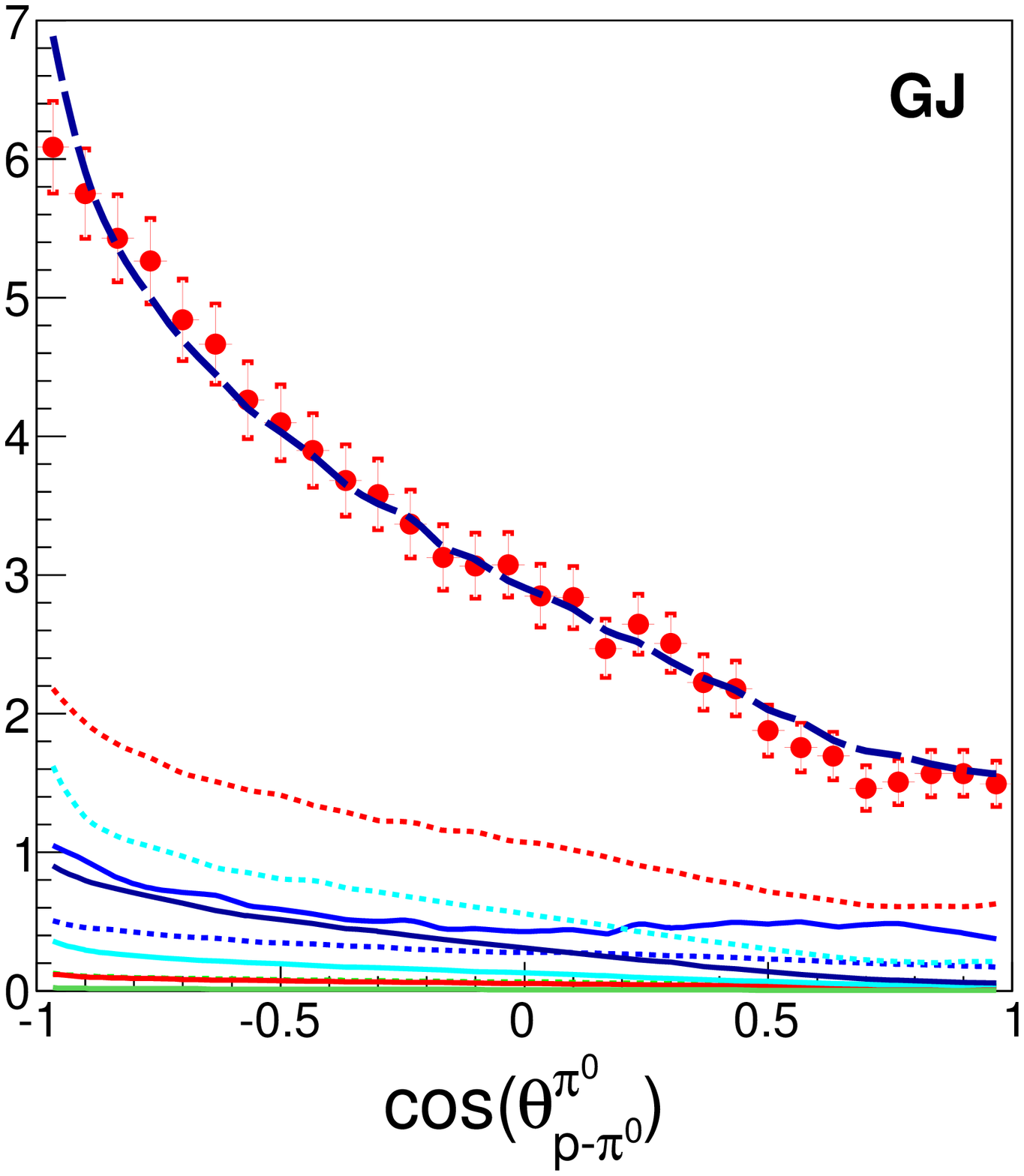}
  \includegraphics{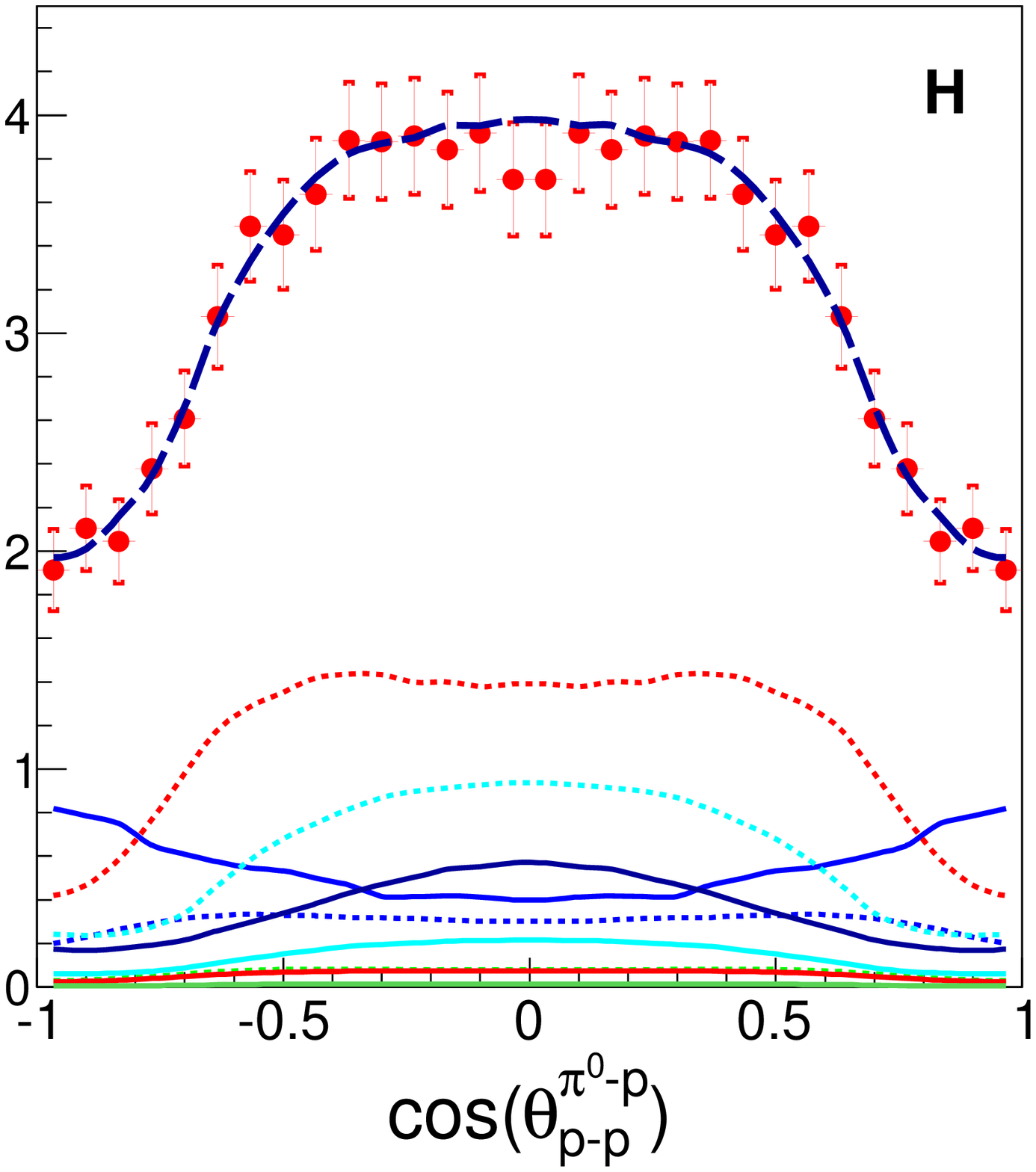}
  \includegraphics{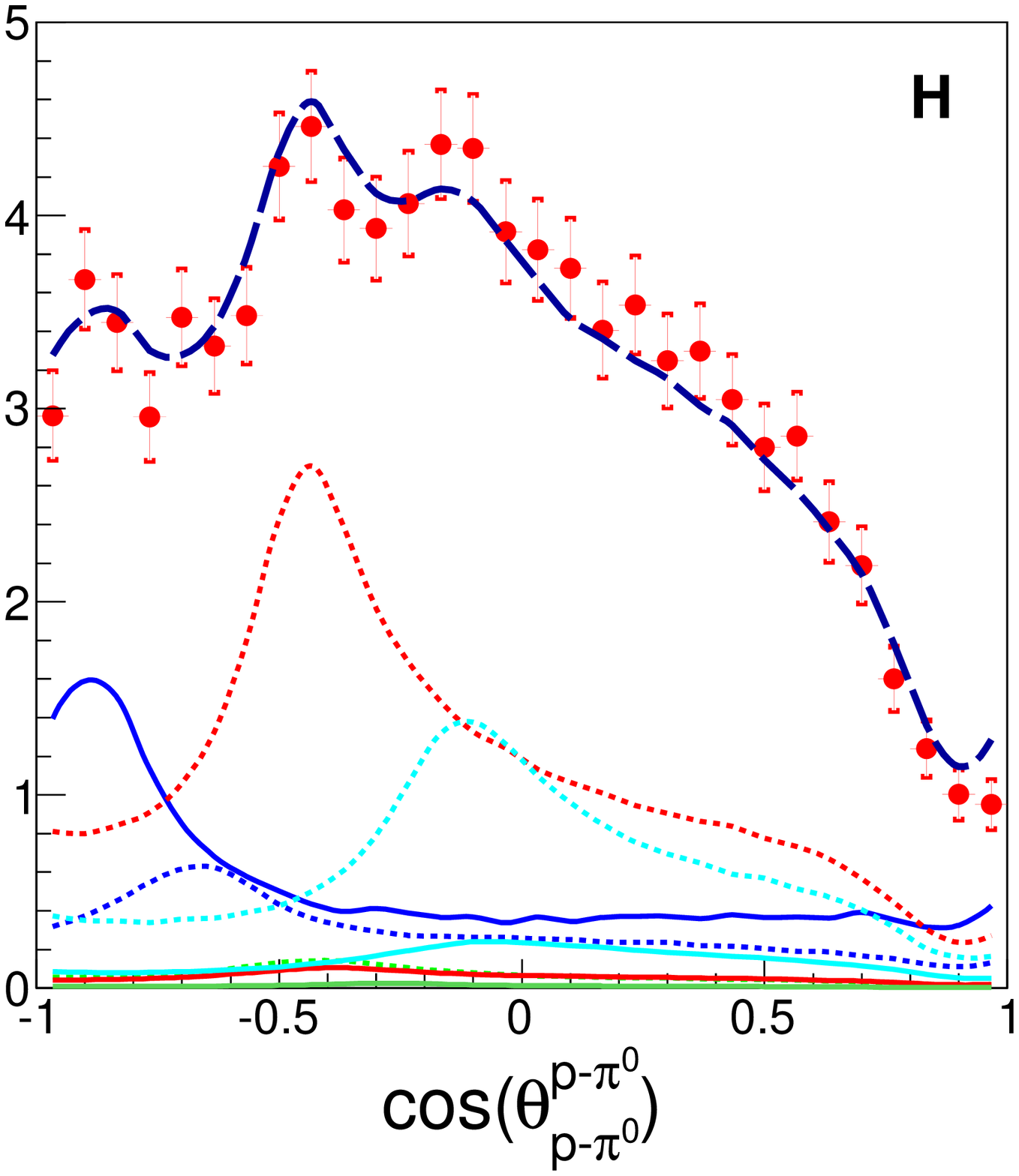}
}
\caption{Angular distributions in the Gottfried-Jackson (GJ) and the helicity (H) reference frames for the $pp\pi^0$ final state compared to the results of simulations (dashed curves) decomposed into contributions of various resonances (line style as in Fig. \ref{pnpi_mass}).}
\label{GJH_ppi}       
\end{figure*}

In order to visualize the good description of the data by our resonance model calculations we present angular distributions in the Gottfried-Jackson ($GJ$) and the helicity ($H$) reference frames. We employ the same notation and definitions of the respective angles as given in our previous work \cite{ppsigma}. For example, in the notation $\theta_{p-\pi^+}^{n-\pi^+}$, the lower label defines the H rest frame of the two particle system ($p-\pi^+$) in which all the momentum vectors are calculated, and the upper label denotes the momentum vectors (in this case, the neutron and pion) used for the opening angle calculation. For the $GJ$ reference frame, only one index is used since the angle is always calculated with respect to the beam particle direction.

Figure \ref{GJH_pnpi1} displays the angular distributions for the $pn\pi^+$ final state in the $GJ$ reference frame and the angular distributions in the $H$ reference frame. Although they are strongly affected by the HADES acceptance they still reveal interesting features related to resonance production. The helicity distributions are connected to the invariant mass distributions and exhibit structures which related to the contributions of individual resonances. As expected, the $n\pi^+$ helicity frame allows to reveal the $p\pi^+$ states. In the case of  $p\pi^+$ helicity frame, the resonant states  deriving from the single charge states, are covered by the decay pattern of the $\Delta^{++}$ resonances.

The angular distributions of nucleons calculated in the $GJ$ frame display a strong forward-backward peaking. The angle $\theta_{p\pi^+}^{\pi^+}$ in the $GJ$ frame describes the decay angle of the double-charged $\Delta^{++}$ and should be sensitive to the expected anisotropy of the $\Delta(1232)$ decay. Indeed, the data seems to follow the trend expected for the $\Delta(1232)$ but are not perfectly described by our simulation. This might be a consequence of the isotropically modeled decays of the other resonances. However, we found only a small sensitivity to modeling of these distributions within the HADES acceptance.

Figure~\ref{GJH_ppi} displays the angular distributions in the $GJ$ and $H$ reference frames for the $pp\pi^0$ final state. The same definitions of angles and notations are used as for the $pn\pi^+$ final state. Since the final state includes two indistinguishable protons only four distributions are presented. The two distributions including two protons were averaged, as explained above. As one can see, for the $pp\pi^0$ reaction even a better description of the data by our model has been achieved. It is interesting to note that the $GJ$ distribution for the $p\pi^0$ system, which is dominated by the $N^*$ contributions (particularly $N(1520)$), is well described by our simulations, hence corroborating our assumption of an isotropic resonance decay.

\end{document}